\documentstyle[axodraw,epsf]{article}

\newcommand{\dslash}[1]{\not\!{#1}}

\begin{document}
\title{
Supersymmetric electroweak corrections to $t \bar{t} h$ associated production
at $e^+e^-$ colliders
}

\author{
Xiao-Hong Wu$^{1,2}$\footnote{Email: wuxh@th.phy.pku.edu.cn},
Chong Sheng Li$^1$\footnote{Email: csli@pku.edu.cn}, and
Jian Jun Liu$^1$\footnote{Email: jjliu@pku.edu.cn}\\
$^1$Department of Physics, Peking University, Beijing 100871, P. R. China \\
$^2$ CCAST (World Lab.), P.O. Box 8730, Beijing 100080, China
}

\date{August 1, 2003}

\maketitle
\begin{abstract}
The electroweak corrections of order ${\cal O}(\alpha_{\rm em}
m^2_{t,b}/m^2_w)$ and ${\cal O}(\alpha_{\rm em} m^3_{t,b}/m^3_w)$
to Higgs boson associated production with top quark pair are
calculated at $e^+e^-$ colliders in the standard model (SM), the
two-Higgs-doublet model (2HDM) and the minimal supersymmetric
standard model (MSSM). These corrections are a few percent in
general, and in the MSSM they can be over ten percent for
favorable parameter values allowing by current precise
experiments. The total cross sections including the electroweak
corrections get their maximal near $\sqrt{s} = 700$GeV, and can
reach 2.8 fb, 2.7 fb and 2.5 fb in the SM, the 2HDM and the MSSM,
respectively.
\end{abstract}


\newpage
\section{Introduction}
Higgs mechanism~\cite{higgs} is a milestone to the establishment
of Standard Model~\cite{sm} (SM), where the $SU(2)_L$ Higgs
doublet generates electroweak symmetry breaking, gives masses to
the weak gauge bosons $W^\pm$, $Z$ through the kinematic terms of
Higgs fields, and gives masses to fermions through Yukawa coupling
of Higgs with fermions. In the SM with one $SU(2)_L$ Higgs
doublet, there is only one physical scalar Higgs boson left after
the electroweak symmetry breaking, the CP even Higgs boson $h^{\rm
sm}$, while in the two-Higgs-doublet model (2HDM) with two
$SU(2)_L$ Higgs doublets, there are three neutral and two charged
Higgs bosons $h$, $H$, $A$, and $H^\pm$, of which $h$ and $H$ are
$CP$ even and $A$ is $CP$ odd. Due to a fundamental scalar $h^{\rm
sm}$, there exists "hierarchy problem" in the SM, where the Higgs
self-energy is quadratic divergent from loop of Dirac fermions,
which pulls the Higgs mass to a high cut-off
scale~\cite{hierarchy}. Supersymmetry is introduced to solve this
problem, through the cancellation of quadratic divergence in the
loops of Higgs self-energy, in which 2HDM of usual type II have to
be introduced~\cite{higgshunter}, where one Higgs doublet couples
to weak isospin $I = \frac{1}{2}$ fermions, and the other Higgs
doublet couples to $I = - \frac{1}{2}$ fermions.

The Search for the Higgs boson is one of major objectives of
future high-energy collider, after the discovery of top
quark~\cite{topfound}. Great effort has been devoted to this
direction to study different properties of the Higgs boson. The
direct search of Higgs boson at LEP II through the process $e^+e^-
\rightarrow Z h^{\rm sm}$ has set the lower experimental limit of
$114.1$GeV on the Standard Model CP even Higgs boson
mass~\cite{lep2higgssm}. Indirect constraint on the mass of Higgs
boson from the W mass measured at Tevatron and LEP II and top
quark mass measured at Tevatron is below $195$GeV at $95\%$
confidence level~\cite{lepewwg02}, which signifies a light Higgs
boson. The latest lower limit on the lightest CP-even Higgs boson
mass in MSSM is $m_h\ge 91.0$GeV ~\cite{lep2higgsmssm,sopczak},
where production process $e^+e^- \rightarrow Z h$ is suppressed
with small $\sin^2(\beta - \alpha)$. When $\sin^2(\beta - \alpha)$
is not suppressed, the limit on the lightest CP-even Higgs boson
mass from above process is about $114$GeV~\cite{sopczak} in the
MSSM. While in the MSSM, because supersymmetry relates Higgs
quadratic coupling to the gauge coupling, there exists an upper
limit on the lightest CP-even Higgs boson mass $m_h \le
130$GeV~\cite{hmax} when including radiative corrections.

With a heavy top quark, $t \bar{t} h$ associated production also
can be helpful for both the discovery of a Higgs boson in the
intermediate mass range and the measurement of large top quark
Yukawa coupling at the near future colliders. There are the
possibilities at the Tevatron Run II to search the Higgs boson
through $q \bar{q}\rightarrow t \bar{t} h$ production subprocess,
followed by $h \rightarrow b \bar{b}$, for $m_h \le
140$GeV~\cite{tevatrontth}, though the statistics are too low to
measure $t \bar{t} h$ Yukawa coupling. At the CERN Large Hadron
Collider (LHC), the process $pp \rightarrow t \bar{t} h$ is an
important search channel for a light Higgs mass from $100$GeV to
$130$GeV~\cite{lhctth}, and the statistics will allow a
measurement of $t \bar{t} h$ Yukawa coupling at $11.9\%$ level
with $m_h = 120$GeV~\cite{lhcyukawa}. Recently, at the Tevatron
and the LHC, ${\cal O}(\alpha_s)$ QCD corrections to $t \bar{t} h$
production processes have been investigated in
Ref.~\cite{hadronqcdzerwas,hadronqcddawson}. At $e^+e^-$
colliders, the measurement of the $t \bar{t} h$ Yukawa coupling
can be significantly improved through the process $e^+e^-
\rightarrow t \bar{t} h$. With integrated luminosity of $1000{\rm
fb}^{-1}$, at $\sqrt{s} = 500$GeV, the top Yukawa coupling can be
measured at an accuracy of $\delta g_{tth}/g_{tth} \simeq 21\%$,
and with integrated luminosity of $500{\rm fb}^{-1}$, at $\sqrt{s}
= 800$GeV, the measurement of top Yukawa coupling can reach the
level of $\delta g_{tth}/g_{tth} \simeq 5\%$~\cite{lctth}. As
shown in Ref.~\cite{lcqcd} that since the $t \bar{t} h$ Yukawa
coupling can be significantly different in the supersymmetric
(SUSY) model from one in the SM, the measurement would provide a
mean of discriminating between models. Thus the theoretical
predictions including higher-order quantum effects in the
different models, which also enters into the rate and dilutes the
interpretation of the signal as the measurement of the $t \bar{t}
h$ coupling, should be important. At $e^+e^-$ linear colliders, $t
\bar{t} h$ associated production process was discussed at
tree-level many years ago~\cite{treetth}, and its QCD corrections
were also given in Ref.~\cite{lcqcd}. Very recently, a calculation
of the SUSY-QCD corrections to the process has been presented in
Ref.~\cite{shzhu}. In this paper, we present the calculation of
the  ${\cal O}(\alpha_{\rm em} m^2_{t,b}/m^2_w)$ and ${\cal
O}(\alpha_{\rm em} m^3_{t,b}/m^3_w)$ electroweak corrections in
the SM, the 2HDM of type II, and the MSSM.

The paper is arranged as follow. In section \ref{sect:anacalc},
we present the notation and analytical calculations.
In section \ref{sect:numerical},
we give our numerical results and discussions.
our conclusions are given in section \ref{sect:conclusion}.
The lengthy explicit expressions of the irreducible self-energies
and vertices are given in the Appendix.

\section{Evaluation}\label{sect:anacalc}

\subsection{Formalism}
Firstly, we define the kinematical variables and list a compact
formula for the one-loop virtual corrections to $e^-e^+
\rightarrow t\bar{t}h$.  The momenta of incoming electron $e^-$
and positron $e^+$ are $p_1$ and $p_2$, respectively, and the
momenta of outgoing top quark $t$, anti-top quark $\bar{t}$ and
the Higgs boson $h$ are assigned to $k_1$, $k_2$ and $k_3$
correspondingly, with the center of mass energy squared $s = (p_1
+ p_2)^2 = (k_1 + k_2 + k_3)^2$, for the process
\begin{equation}
e^-(p_1) + e^+(p_2) \rightarrow t(k_1) + \bar{t}(k_2) + h(k_3).
\label{eetth}
\end{equation}
Because of the smallness of electron mass, we neglect
Higgs-electron-electron Yukawa coupling, the $2 \rightarrow 3$
process (\ref{eetth}) can be split to two parts: the electron and
positron are annihilated to time-like $\gamma$ and $Z$, and then
the virtual $\gamma$, $Z$ decay to $t \bar{t} h$ final states.
Therefore, the differential cross section of the process
(\ref{eetth}) with averaged over spin of incoming states and
summed over spin and color of outgoing states can be simplified,
which should only depend on two parameters, conveniently chosen as
the energy of outgoing top quark $k^0_1$, and the energy of
outgoing Higgs boson $k^0_3$, and the differential cross section
including the one-loop virtual corrections can be written in the
general form
\begin{eqnarray}
\frac{d\sigma}{d k^0_3 d k^0_1} &=& \frac{N_c}{4} \frac{1}{8 (2\pi)^4}
 \frac{1}{2 s}
\overline{\sum_{\rm spins}}[|M^{\rm tree}|^2 + 2 {\rm Re}(M^{\rm
tree} \delta M^+) ],
\end{eqnarray}
where $N_c$ is the $SU(3)_c$ color factor, $M^{\rm tree}$ is tree
level amplitude, and $\delta M$ is the one-loop corrections to the
tree-level process, which contain the irreducible corrections
$\delta M^{({\rm self, vertex, box})}_{\rm irr}$ and the corresponding
conterterms $\delta M^{({\rm self, vertex})}_{\rm ct}$:

\begin{eqnarray}
\delta M &=& \delta M^{\rm self} + \delta M^{\rm vertex} +
 \delta M^{\rm box}_{\rm irr}
\end{eqnarray}

with

\begin{eqnarray}
\delta M^{\rm self} &=& \delta M^{\rm self}_{\rm irr} +
 \delta M^{\rm self}_{\rm ct}, \nonumber\\
\delta M^{\rm vertex} &=& \delta M^{\rm vertex}_{\rm irr} +
 \delta M^{\rm vertex}_{\rm ct}.
\end{eqnarray}

\subsection{SUSY electroweak corrections}
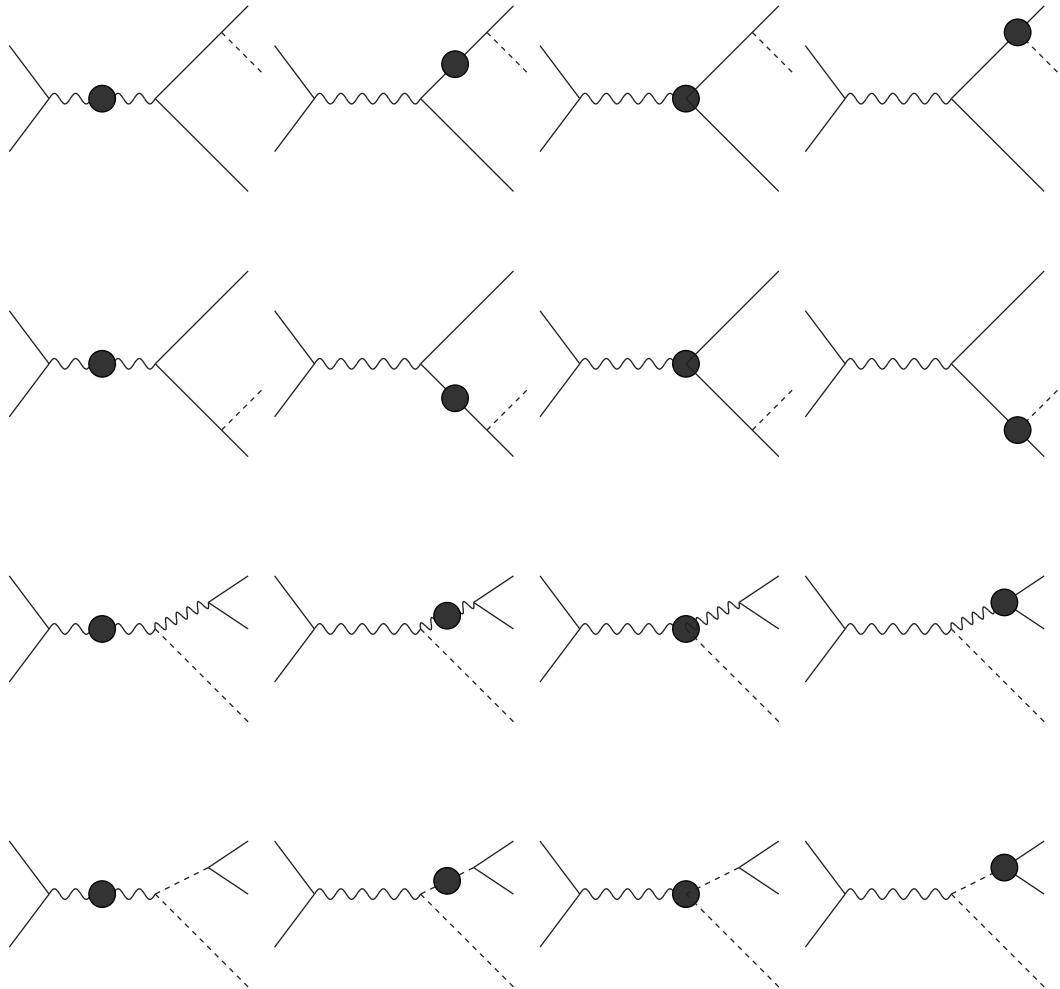
\begin{figure}
\begin{center}
\begin{picture}(400,400)(0,0)
\Line(5,370)(20,350)
\Line(20,350)(5,330)
\Photon(20,350)(60,350){2}{5}
\GCirc(40,350){5}{0.2}
\Line(60,350)(95,385)
\DashLine(85,375)(100,360){2}
\Line(60,350)(95,315)
\Line(105,370)(120,350)
\Line(120,350)(105,330)
\Photon(120,350)(160,350){2}{5}
\Line(160,350)(195,385)
\GCirc(173,363){5}{0.2}
\DashLine(185,375)(200,360){2}
\Line(160,350)(195,315)
\Line(205,370)(220,350)
\Line(220,350)(205,330)
\Photon(220,350)(260,350){2}{5}
\GCirc(260,350){5}{0.2}
\Line(260,350)(295,385)
\DashLine(285,375)(300,360){2}
\Line(260,350)(295,315)
\Line(305,370)(320,350)
\Line(320,350)(305,330)
\Photon(320,350)(360,350){2}{5}
\Line(360,350)(395,385)
\DashLine(385,375)(400,360){2}
\GCirc(385,375){5}{0.2}
\Line(360,350)(395,315)
\Line(5,270)(20,250)
\Line(20,250)(5,230)
\Photon(20,250)(60,250){2}{5}
\GCirc(40,250){5}{0.2}
\Line(60,250)(95,285)
\Line(60,250)(95,215)
\DashLine(85,225)(100,240){2}
\Line(105,270)(120,250)
\Line(120,250)(105,230)
\Photon(120,250)(160,250){2}{5}
\Line(160,250)(195,285)
\Line(160,250)(195,215)
\GCirc(173,237){5}{0.2}
\DashLine(185,225)(200,240){2}
\Line(205,270)(220,250)
\Line(220,250)(205,230)
\Photon(220,250)(260,250){2}{5}
\GCirc(260,250){5}{0.2}
\Line(260,250)(295,285)
\Line(260,250)(295,215)
\DashLine(285,225)(300,240){2}
\Line(305,270)(320,250)
\Line(320,250)(305,230)
\Photon(320,250)(360,250){2}{5}
\Line(360,250)(395,285)
\Line(360,250)(395,215)
\DashLine(385,225)(400,240){2}
\GCirc(385,225){5}{0.2}
\Line(5,170)(20,150)
\Line(20,150)(5,130)
\Photon(20,150)(60,150){2}{5}
\GCirc(40,150){5}{0.2}
\Photon(60,150)(80,160){2}{5}
\Line(80,160)(95,170)
\Line(80,160)(95,150)
\DashLine(60,150)(95,115){2}
\Line(105,170)(120,150)
\Line(120,150)(105,130)
\Photon(120,150)(160,150){2}{5}
\Photon(160,150)(180,160){2}{5}
\GCirc(170,155){5}{0.2}
\Line(180,160)(195,170)
\Line(180,160)(195,150)
\DashLine(160,150)(195,115){2}
\Line(205,170)(220,150)
\Line(220,150)(205,130)
\Photon(220,150)(260,150){2}{5}
\GCirc(260,150){5}{0.2}
\Photon(260,150)(280,160){2}{5}
\Line(280,160)(295,170)
\Line(280,160)(295,150)
\DashLine(260,150)(295,115){2}
\Line(305,170)(320,150)
\Line(320,150)(305,130)
\Photon(320,150)(360,150){2}{5}
\Photon(360,150)(380,160){2}{5}
\Line(380,160)(395,170)
\Line(380,160)(395,150)
\GCirc(380,160){5}{0.2}
\DashLine(360,150)(395,115){2}
\Line(5,70)(20,50)
\Line(20,50)(5,30)
\Photon(20,50)(60,50){2}{5}
\GCirc(40,50){5}{0.2}
\DashLine(60,50)(80,60){2}
\Line(80,60)(95,70)
\Line(80,60)(95,50)
\DashLine(60,50)(95,15){2}
\Line(105,70)(120,50)
\Line(120,50)(105,30)
\Photon(120,50)(160,50){2}{5}
\DashLine(160,50)(180,60){2}
\GCirc(170,55){5}{0.2}
\Line(180,60)(195,70)
\Line(180,60)(195,50)
\DashLine(160,50)(195,15){2}
\Line(205,70)(220,50)
\Line(220,50)(205,30)
\Photon(220,50)(260,50){2}{5}
\GCirc(260,50){5}{0.2}
\DashLine(260,50)(280,60){2}
\Line(280,60)(295,70)
\Line(280,60)(295,50)
\DashLine(260,50)(295,15){2}
\Line(305,70)(320,50)
\Line(320,50)(305,30)
\Photon(320,50)(360,50){2}{5}
\DashLine(360,50)(380,60){2}
\Line(380,60)(395,70)
\Line(380,60)(395,50)
\GCirc(380,60){5}{0.2}
\DashLine(360,50)(395,15){2}
\end{picture}
\end{center}
\caption[]{One-loop corrected diagrams of process $e^+ e^-
\rightarrow t \bar{t} h$.} \label{oneloop}
\end{figure}

The SUSY electroweak corrections of order ${\cal O}(\alpha_{\rm
em} m^2_{t,b}/m^2_w)$ and ${\cal O}(\alpha_{\rm em}
m^3_{t,b}/m^3_w)$ to process (\ref{eetth}) arise from the Feynman
diagrams shown in {\bf Fig}.~\ref{oneloop}. We carried out the
calculation in the 't Hooft-Feynman gauge and used dimensional
reduction to control all the ultraviolate divergences in the
virtual loop corrections utilizing the on-mass-shell
renormalization scheme~\cite{onshell,denner}.
FeynCalc~\cite{feyncalc} is used to calculate the one-loop
irreducible diagrams.

The relevant field renormalization constants are defined as
\begin{eqnarray}
t_{L0} &=& (1 + \frac{1}{2} \delta Z^t_L) t_L,
\hspace{5mm} t_{R0} = (1 + \frac{1}{2} \delta Z^t_R) t_R, \nonumber\\
\left(
\begin{array}{c}
Z_\mu \\ A_\mu
\end{array}
\right)_0 &=&
\left(
\begin{array}{cc}
1 + \frac{1}{2} \delta Z_{zz} & \frac{1}{2} \delta Z_{z\gamma} \\
\frac{1}{2} \delta Z_{\gamma z} & 1 + \frac{1}{2} \delta Z_{\gamma\gamma}
\end{array}
\right)
\left(
\begin{array}{c}
Z_\mu \\ A_\mu
\end{array}
\right), \nonumber\\
\left(
\begin{array}{c}
H \\ h
\end{array}
\right)_0 &=&
\left(
\begin{array}{cc}
1 + \frac{1}{2} \delta Z_{HH} & \frac{1}{2} \delta Z_{Hh} \\
\frac{1}{2} \delta Z_{hH} & 1 + \frac{1}{2} \delta Z_{hh}
\end{array}
\right)
\left(
\begin{array}{c}
H \\ h
\end{array}
\right), \nonumber\\
\left(
\begin{array}{c}
A \\ G
\end{array}
\right)_0 &=&
\left(
\begin{array}{cc}
1 + \frac{1}{2} \delta Z_{AA} & \frac{1}{2} \delta Z_{AG} \\
\frac{1}{2} \delta Z_{GA} & 1 + \frac{1}{2} \delta Z_{GG}
\end{array}
\right)
 \left(
\begin{array}{c}
A \\ G
\end{array}
\right).
\end{eqnarray}
The mass renormalization constants are defined as
\begin{eqnarray}
m^2_{w0} &=& m^2_w + \delta m^2_w,
\hspace{5mm} m^2_{z0} = m^2_z + \delta m^2_z, \nonumber\\
m_{t0} &=& m_t + \delta m_t, \nonumber\\
m^2_{h0} &=& m^2_h + \delta m^2_h, \nonumber\\
m^2_{H0} &=& m^2_H + \delta m^2_H, \nonumber\\
m^2_{A0} &=& m^2_A + \delta m^2_A.
\end{eqnarray}
As for the CP-even Higgs mixing angle $\alpha$ and $\tan\beta
\equiv \upsilon_2/\upsilon_1$, the ratio of two vacuum expectation
values of neutral Higgs fields, they have to be renormalized, too.
We defined
\begin{eqnarray}
\alpha_0 &=& (1 + \delta Z_\alpha) \alpha, \nonumber\\
(\tan\beta)_0&=& (1 + \delta Z_\beta) \tan\beta,
\end{eqnarray}
where $\alpha$ and $\beta$ are not independent in the MSSM.

Below we described in detail the renormalization constants and
counterterms in the 2HDM and the MSSM, while ones in the SM can be
obtained using standard techniques~\cite{denner}. For the Higgs
sector of the MSSM, using the results of Ref.~\cite{santos}, the
relevant renormalization constants are given by
\begin{eqnarray}
\delta m_h^2 &=& \Sigma^{hh}(m_h^2) - T_{hh}, \nonumber\\
\delta m_H^2 &=& \Sigma^{HH}(m_H^2) - T_{HH}, \nonumber\\
\delta Z_{hh} &=& - \frac{\partial \Sigma^{hh}(k^2)}{\partial
k^2}|_{k^2=m_h^2},
\nonumber\\
\delta Z_{HH} &=& - \frac{\partial \Sigma^{HH}(k^2)}{\partial
k^2}|_{k^2=m_H^2},
\nonumber\\
\delta Z_{hH} &=& -2 \frac{\Sigma^{Hh}(m_H^2) - T_{Hh}}{m_H^2 -
m_h^2},
\nonumber\\
\delta Z_{Hh} &=& 2 \frac{\Sigma^{Hh}(m_h^2) - T_{Hh}}{m_H^2 -
m_h^2},
\nonumber\\
\delta m_A &=& \Sigma^{AA}(m_A^2) - T_{AA}, \nonumber\\
\delta Z_{AA} &=& - \frac{\partial \Sigma^{AA}(k^2)}{\partial
k^2}|_{k^2=m_A^2},
\nonumber\\
\delta Z_{GG} &=& - \frac{\partial \Sigma^{GG}(k^2)}{\partial
k^2}|_{k^2=0},
\nonumber\\
\delta Z_{GA} &=& -2 \frac{\Sigma^{AG}(m_A^2) - T_{AG}}{m_A^2},
\nonumber\\
\delta Z_{AG} &=& 2 \frac{\Sigma^{AG}(0) - T_{AG}}{m_A^2},
\end{eqnarray}
where the Higgs tadpole parameters $T_h$, $T_H$, $T_{HH}$,
$T_{Hh}$, $T_{hh}$, $T_{AA}$, $T_{AG}$, $T_{GG}$, $T_{H^-H^-}$,
$T_{H^-G^-}$, and $T_{G^-G^-}$ defined as in Ref.~\cite{santos}.
The renormalization of CP-even Higgs mixing angle $\alpha$ is
fixed by ~\cite{hollik}
\begin{eqnarray}
\delta Z_\alpha &=& \frac{1}{4} (\delta Z_{Hh} - \delta Z_{hH}).
\end{eqnarray}
The renormalization of $\tan\beta$ is fixed by keeping the
on-shell $H^+ \bar{\tau} \nu_\tau$ coupling the same form to all
orders in perturbative series~\cite{tanb}, which can be expressed
as
\begin{eqnarray}
\delta Z_\beta &=& \frac{1}{2} \tan\beta ( \frac{\delta m_w^2}{m_w^2}
 - \frac{\delta m_z^2}{m_z^2} +
 \frac{\delta m_z^2 - \delta m_w^2}{m_z^2 - m_w^2} - \delta Z_{H^-H^-} +
 \delta Z_{G^-H^-} \cot\beta ) \nonumber\\
\end{eqnarray}
with
\begin{eqnarray}
\delta Z_{H^-H^-} &=& - \frac{\partial
\Sigma^{H^-H^-}(k^2)}{\partial k^2}|_{k^2=m_{H^-}^2},
\nonumber\\
\delta Z_{G^-H^-} &=& -2 \frac{\Sigma^{H^-G^-}(m_{H^-}^2) -
 T_{H^-G^-}}{m_{H^-}^2}.
\end{eqnarray}

Using above the renormalization constants, for the process
(\ref{eetth}), the counterterms of Higgs self-energy can be
expressed as
\begin{eqnarray}
C_{hh} &=& i [k^2 \delta Z_{hh} - m_h^2 \delta Z_{hh} - \delta
m_h^2 + T_{hh}],
\nonumber\\
C_{HH} &=& i [k^2 \delta Z_{HH} - m_H^2 \delta Z_{HH} - \delta
m_H^2 + T_{HH}],
\nonumber\\
C_{hH} &=& i [\frac{1}{2}(\delta Z_{hH} + \delta Z_{Hh}) k^2 -
 \frac{1}{2}(m_h^2 \delta Z_{hH} + m_H^2 \delta Z_{Hh}) + T_{Hh}], \nonumber\\
C_{GG} &=& i [\delta Z_{GG} k^2 + T_{GG}], \nonumber\\
C_{AA} &=& i [k^2 \delta Z_{AA} - m_A^2 \delta Z_{AA} - \delta
m_A^2 + T_{AA}],
\nonumber\\
C_{AG} &=& i [\frac{1}{2}(\delta Z_{AG} + \delta Z_{GA}) k^2 -
 \frac{1}{2} m_A^2 \delta Z_{AG} + T_{AG}].
\end{eqnarray}

The counterterms of gauge boson and CP-odd scalar mixing, $Z_\mu -
G$ and $Z_\mu - A$ are  given by
\begin{eqnarray}
C_{zG} &=& k_\mu m_z [\frac{\delta m_z^2}{2 m_z^2} +
 \frac{\delta Z_{zz}}{2} + \frac{\delta Z_{GG}}{2}], \nonumber\\
C_{zA} &=& k_\mu m_z \frac{\delta Z_{GA}}{2}.
\end{eqnarray}

The counterterms of gauge boson and top pair interactions, $\gamma
t \bar{t}$ and $Z t \bar{t}$ are given by
\begin{eqnarray}
C_{\gamma tt} &=& [g^L_{tt\gamma} (\frac{\delta Z_{\gamma\gamma}}{2} +
 \delta Z^t_L) +
 g^L_{ttz} \frac{\delta Z_{z\gamma}}{2} + g^L_{tt\gamma} \frac{\delta e}{e}]
 \gamma_\mu P_L + (L \rightarrow R), \nonumber\\
C_{ztt} &=& [g^L_{ttz} (\frac{\delta Z_{zz}}{2} + \delta Z^t_L) +
 g^L_{tt\gamma} \frac{\delta Z_{\gamma z}}{2} +
 g^L_{ttz} (\frac{\delta e}{e} +
 \frac{8 s_w \delta s_w}{4 s_w^2 - 3} - \frac{\delta s_w}{s_w} -
 \frac{\delta c_w}{c_w} ] \gamma_\mu P_L + (L \rightarrow R),
\end{eqnarray}
where $P_{L,R} = (1 \mp \gamma_5)/2$, and $\delta e/e$ is absent
at the order of ${\cal O}({\alpha_{\rm em} m_{t,b}^2/m_w^2})$
corrections.

The counterterms of Higgs and top pair Yukawa interactions, $h t
\bar{t}$, $G t \bar{t}$ and $A t \bar{t}$ are given by
\begin{eqnarray}
C_{htt} &=& [g^L_{tth} (\frac{\delta Z_{hh}}{2} + \frac{\delta Z^L_t}{2} +
 \frac{\delta Z^R_t}{2}) +
 g^L_{tth} (\frac{\delta m_t}{m_t} + \frac{\delta e}{e} -
 \frac{\delta m_w^2}{2 m_w^2} - \frac{\delta s_w}{s_w} -
 \frac{\delta\sin\beta}{\sin\beta} +
 \frac{\delta\cos\alpha}{\cos\alpha}) + \nonumber\\
&& g^L_{ttH} \frac{\delta Z_{Hh}}{2}] P_L + (L \rightarrow R), \nonumber\\
C_{Gtt} &=& [g^L_{ttG} (\frac{\delta Z_{GG}}{2} + \frac{\delta Z^L_t}{2} +
 \frac{\delta Z^R_t}{2}) +
 g^L_{ttG} (\frac{\delta m_t}{m_t} + \frac{\delta e}{e} -
 \frac{\delta m_w^2}{2 m_w^2} - \frac{\delta s_w}{s_w} ) + \nonumber\\
&& g^L_{ttA} \frac{\delta Z_{AG}}{2}] P_L + (L \rightarrow R), \nonumber\\
C_{Att} &=& [g^L_{ttA} (\frac{\delta Z_{AA}}{2} + \frac{\delta Z^L_t}{2} +
 \frac{\delta Z^R_t}{2}) +
 g^L_{ttA} (\frac{\delta m_t}{m_t} + \frac{\delta e}{e} -
 \frac{\delta m_w^2}{2 m_w^2} - \frac{\delta s_w}{s_w} -
 \frac{\delta\cot\beta}{\cot\beta}) + \nonumber\\
&& g^L_{ttG} \frac{\delta Z_{GA}}{2}] P_L + (L \rightarrow R).
\end{eqnarray}

Note that $g^L_{tti}$ ($i=\gamma$, H, h, A, G, Z) appeared above
are the coupling constants.

The counterterms of Higgs and gauge boson pair interactions,
$h\gamma\gamma$, $h\gamma z$ and $h z z$ are given by
\begin{eqnarray}
C_{h\gamma\gamma} &=& 0, \nonumber\\
C_{h\gamma z} &=& g_{zzh} \frac{\delta Z_{z\gamma}}{2} g_{\mu\nu}, \nonumber\\
C_{hzz} &=& [g_{zzh} (\frac{\delta Z_{hh}}{2} + \delta Z_{zz}) +
 g_{zzh} (\frac{\delta g}{g} + \frac{\delta m_w^2}{2 m_w^2} -
 2 \frac{\delta c_w}{c_w} +
 \frac{\delta \sin (\beta-\alpha)}{\sin (\beta-\alpha)}) +
 g_{zzH} \frac{\delta Z_{Hh}}{2}] g_{\mu\nu}.
\end{eqnarray}

The counterterms of gauge boson and Higgs pair interactions,
$\gamma h G$, $\gamma h A$, $Z h G$ and $Z h A$ are given by
\begin{eqnarray}
C_{\gamma hG} &=& g_{hGz} \frac{\delta Z_{z\gamma}}{2} (k_h -
k_G)_\mu,
 \nonumber\\
C_{\gamma hA} &=& g_{hAz} \frac{\delta Z_{z\gamma}}{2} (k_h -
k_A)_\mu,
 \nonumber\\
C_{zhG} &=& [g_{hGz} (\frac{\delta Z_{zz}}{2} + \frac{\delta Z_{GG}}{2} +
 \frac{\delta Z_{hh}}{2}) + g_{hGz} (\frac{\delta e}{e} -
 \frac{\delta c_w}{c_w} - \frac{\delta s_w}{s_w} +
 \frac{\delta \sin(\beta-\alpha)}{\sin(\beta-\alpha)}) + \nonumber\\
&& g_{hAz} \frac{\delta Z_{AG}}{2} +
 g_{HGz} \frac{\delta Z_{Hh}}{2}] (k_h - k_G)_\mu, \nonumber\\
C_{zhA} &=& [g_{hAz} (\frac{\delta Z_{zz}}{2} + \frac{\delta Z_{AA}}{2} +
 \frac{\delta Z_{hh}}{2}) + g_{hAz} (\frac{\delta e}{e} -
 \frac{\delta c_w}{c_w} - \frac{\delta s_w}{s_w} +
 \frac{\delta \cos(\beta-\alpha)}{\cos(\beta-\alpha)}) + \nonumber\\
&& g_{hGz} \frac{\delta Z_{GA}}{2} +
 g_{HAz} \frac{\delta Z_{Hh}}{2}] (k_h - k_A)_\mu.
\end{eqnarray}

With all the counterterms fixed as above, the renormalized
amplitude for the process (\ref{eetth}) is obtained by adding the
counterterms to the corresponding irreducible corrections arising
from the self-energy diagrams shown in {\bf Fig}.~\ref{vv}
--~\ref{tt}, the vertex diagrams shown in {\bf Fig}.~\ref{vtt}
--~\ref{vss}, and the box diagrams shown in {\bf Fig}.~\ref{box},
respectively, which can be reduced by FeynCalc~\cite{feyncalc} and
the relevant explicit expressions are given in the Appendix. Using
above results it is straightforward to calculate the ${\cal
O}(\alpha_{\rm em} m^2_{t,b}/m^2_w)$ and ${\cal O}(\alpha_{\rm em}
m^3_{t,b}/m^3_w)$ SUSY electroweak corrections to process
(\ref{eetth}).

\section{Numerical results and Discussion}\label{sect:numerical}
In this section we present some numerical results. In the SM and
the 2HDM, we limit the lightest CP-even Higgs boson mass $m_h$
larger than $114$GeV, while in the 2HDM of type II, the lower
bound of the charged Higgs boson mass is $350$GeV from the
constraint of Br($B\rightarrow X_s \gamma$)~\cite{hewett}. In the
MSSM, we use SUBHPOLE~\cite{subhpole} to calculate the radiative
corrections to the CP-even Higgs pole masses and mixing angle
$\alpha$, and the CP-odd Higgs pole mass, which incorporates the
one-loop effective potential and two-loop leading-log
contributions from arbitrary off-diagonal stop and sbottom
matrices. The inclusion of the radiative corrections in the Higgs
sector is essential for Higgs phenomenology, because the
tree-level lightest CP-even Higgs mass $m_h < m_z$ has been
excluded by LEP experiments. There is $\pm 3$GeV inaccuracy in
SUBHPOLE to calculate the lightest CP-even Higgs boson mass, due
to unclear subleading-log and uncalculated higher order
corrections, so we take the lower limit of $m_h$ as $111$GeV. We
also take the lower bound of other SUSY particle masses as,
$m_{\tilde{\chi}^0_1} \ge 37$GeV, $m_{\tilde{\chi}^\pm_1} \ge
94$GeV, $m_{\tilde{t}_1} \ge 95$GeV, $m_{\tilde{b}_1} \ge
91$GeV~\cite{pdg}, and $\Delta\rho \le 3 \times
10^{-3}$~\cite{rho}. In the MSSM, for simplicity, we assume $A_t =
A_b$ and use the relation of gaugino masses $M_2 = 2 M_1$ at the
electroweak scale. In all the three different models, the SM, the
2HDM and the MSSM, there are common features that the contribution
of $\gamma$ exchange channel is dominant in the center of mass
energy $\sqrt{s}$ region from $0.5$TeV to $1$TeV, while the
contribution of $Z$ exchange channel increase with increasing
$\sqrt{s}$.

The results in the SM are shown in {\bf Fig}.~\ref{smfig}
and {\bf Fig}.~\ref{sqrts}. We present the
cross section $\sigma$ and SUSY electroweak
corrections to the cross sections relative to the tree-level
values $\Delta\sigma/\sigma^{\rm tree}$
with $\Delta\sigma \equiv \sigma^{\rm all} - \sigma^{\rm tree}$
as a function of the
center of mass energy $\sqrt{s}$ with $m_h = 115$GeV in {\bf Fig}.~\ref{sqrts}.
The corrections are at most a few percent and
decrease with increasing $\sqrt{s}$. {\bf Fig}.~\ref{smfig} gives the
cross section and the corrections as a function of the Higgs
boson mass $m_h$ at $\sqrt{s} = 800$GeV. The corrections increase
as  $m_h$ increase, but still only a few percent. And we can see
that the corrections are positive when $\sqrt{s}$ varies from
$0.5$TeV to $1$TeV for $m_h = 115$GeV in the SM.

{\bf Fig}.~\ref{thdmfig} and {\bf Fig}.~\ref{sqrts} shows the
results in the 2HDM. {\bf Fig}.~\ref{sqrts} gives $\sigma$ and
$\Delta\sigma/\sigma^{\rm tree}$ as a function of $\sqrt{s}$ for
$\tan\beta = 10$ and $\alpha = 0.05$, assuming $m_h = 115$GeV,
$m_H = 250$GeV, $m_A = 300$GeV and $m_{H^\pm} = 350$GeV. The
corrections imply a few percent reduction in the cross sections,
and the magnitude of the corrections firstly increase as
$\sqrt{s}$ increase, then slowly decrease. {\bf
Fig}.~\ref{thdmfig}$a$ shows the dependence of $\sigma$ and
$\Delta\sigma/\sigma^{\rm tree}$ on $m_h$ for $\tan\beta = 10$ and
$\alpha = 0.05$, assuming $m_H = 300$GeV, $m_A = 300$GeV and
$m_{H^\pm} = 300$GeV. The corrections are negative, and their
magnitude can exceed ten percent which increase as $m_h$ increase.
In {\bf Fig}.~\ref{thdmfig}$b$, we present $\sigma$ and
$\Delta\sigma/\sigma^{\rm tree}$ as a function of $\tan\beta$ for
$\alpha = 0.05$, assuming $m_h = 115$GeV, $m_H = 250$GeV, $m_A =
300$GeV and $m_{H^\pm} = 350$GeV. The corrections are about $2\%$
reduction in the cross sections, and not sensitive to the values
of $\tan\beta$.

In {\bf Fig}.~\ref{mssmfig} --~\ref{sqrts}, we give the results in
the MSSM. {\bf Fig}.~\ref{mssmfig}$a$ shows the dependence of
$\sigma$ and $\Delta\sigma/\sigma^{\rm tree}$ on $\mu$ for
$\tan\beta = 44$, assuming $m_A = 308$GeV, $M_1 = 59$GeV,
$A_t=562$GeV and
$m_{\tilde{t}_R}=m_{\tilde{b}_R}=m_{\tilde{q}}=375$GeV. The total
cross section $\sigma$ only slightly depend on the $\mu$, but has
a strong dependence on the c.m. energy $\sqrt{s}$, as shown in
{\bf Fig}.~\ref{mssmfig}$a$ (left); the magnitude of the
corrections $\Delta\sigma/\sigma^{\rm tree}$ is a few percent in
general, and can exceed $10\%$ for $\mu>700$ GeV, as shown in {\bf
Fig}.~\ref{mssmfig}$a$ (right). {\bf Fig}.~\ref{mssmfig}$b$
exhibits $\sigma$ and $\Delta\sigma/\sigma^{\rm tree}$ as a
function of the lightest scalar top quark mass $m_{\tilde{t}_1}$
for $\tan\beta = 4, 10$ and 40, respectively, with $m_A = 250$GeV,
$\mu = 220$GeV, $M_1 = 65$GeV, $A_t=1100$GeV, and
$m_{\tilde{t}_R}=m_{\tilde{b}_R}=1.3m_{\tilde{q}}$. One can see
that the results are not sensitive to the values of $\tan\beta$,
where there is a cut in the curve for $\tan\beta=4$ with
$m_{\tilde{t}_1}$ below $180$GeV  due to the constraint of the
lightest CP-even Higgs mass lower bound as shown in {\bf
Fig}.~\ref{mssmfig}$b$ (left). With $\sqrt{s}=800$GeV, the
corrections decrease the cross sections, and their magnitudes
become smaller with $m_{\tilde{t}_1}$ increasing, but with
$\sqrt{s}=500$GeV, the corrections can increase or decrease the
cross sections depending on $m_{\tilde{t}_1}$, and they are less
than a few percent in general, however, when $m_{\tilde{t}_1}$ is
below $110$GeV, the corrections can be over $-10\%$ with
$\sqrt{s}=800$GeV as shown in {\bf Fig}.~\ref{mssmfig}$b$ (right).
We show $\sigma$ and $\Delta\sigma/\sigma^{\rm tree}$ as a
function of $\sqrt{s}$ in {\bf Fig}.~\ref{sqrts} for $\tan\beta =
6$ and 40, respectively, with $m_A = 160$GeV, $\mu = 220$GeV, $M_1
= 70$GeV, $A_t=1000$GeV, and
$m_{\tilde{t}_R}=m_{\tilde{b}_R}=1.5m_{\tilde{q}}$. The
corrections decrease the cross sections, and their magnitudes
increase as $\sqrt{s}$ increasing, which range between $0 \sim
8\%$ and $4\% \sim 13\%$ for $\tan\beta=6$ and $\tan\beta=40$,
respectively.

\section{Conclusion}\label{sect:conclusion}
To summarize, we have calculated the electroweak corrections of
order ${\cal O}(\alpha_{\rm em} m^2_{t,b}/m^2_w)$ and ${\cal
O}(\alpha_{\rm em} m^3_{t,b}/m^3_w)$ to process $e^+ e^-
\rightarrow t \bar{t} h$ in the SM, the 2HDM and the MSSM,
respectively. These corrections are a few percent in general, and
can be over $10\%$ in the MSSM with the lightest scalar top and
bottom quark mass near the lower bound. The total cross sections
including the electroweak corrections for process $e^+ e^-
\rightarrow t \bar{t} h$ get their maximal near $\sqrt{s} =
700$GeV, and can reach 2.8 fb, 2.7 fb and 2.5 fb in the SM, the
2HDM and the MSSM, respectively.

{\cal Note added:}  While preparing this manuscript three
papers~\cite{ma,belanger,dennertth} appeared where the electroweak
corrections to the same process in the SM are also calculated, the
numerical results of their virtual weak corrections are in
agreement with our results in the case of the SM.

\section*{Acknowledgments}
One of the authors X.-H.W. would like to thank Dr. Shou-Hua Zhu for
his generous discussion about numerical calculations.

\newpage
\appendix
\section*{\bf Appendix}\nonumber
In this Appendix, we present the irreducible self-energy
and vertex contributions.
The convension, Feynman rules, and coupling constants
agree with Ref.~\cite{haberkane}.
In the formulae followed,
we have employed Passarino$-$Veltman one-loop functions
$B_i$, $B_{ij}$, ($i=0,1$)
$C_i$, $C_{ij}$, $C_{ijk}$ ($i,j,k=0,1,2$)~\cite{feyncalc}
and our notations agree with FeynCalc.

\begin{figure}
\begin{center}
\begin{picture}(100,100)(0,0)
\Photon(0,50)(25,50){2}{5}
\ArrowLine(5,55)(20,55)
\Text(10,60)[bl]{$k$}
\Text(0,40)[bl]{$v_a$}
\ArrowArcn(50,50)(25,0,180)
\Text(53,62)[br]{$f_1$}
\ArrowArcn(50,50)(25,180,360)
\Text(53,39)[tr]{$f_2$}
\Photon(75,50)(100,50){2}{5}
\Text(100,40)[br]{$v_b$}
\Text(34,10)[bl]{${\rm v - v}$}
\end{picture}
\end{center}
\caption[]{Gauge boson self-energy ($v_av_b$).}
\label{vv}
\end{figure}
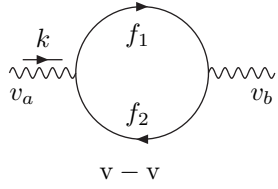
The unrenormalized gauge boson$-$gauge boson $v_a v_b$
($v_a v_b = w^-w^-, \gamma\gamma, \gamma z, zz$) self-energies as shown
in {\bf Fig}.~\ref{vv} are as follow,
which we only extract terms proportional to $m_{t,b}^2$.
\begin{eqnarray}
\Sigma^{\rm v_a v_b} &=& - i \Sigma^{v_a v_b}_T(k^2) g_{\mu\nu} -
 i \Sigma^{v_a v_b}_L(k^2) k_\mu k_\nu \nonumber\\
\Sigma^{ww}_T(k^2) &=& - \frac{g^2 N_c}{48\pi^2}
 [- m_t^2 - m_b^2 + 2 A_0(m_t^2) + m_b^2 B_0(k^2,m_b^2,m_t^2) +
 (m_t^2 - m_b^2) B_1(k^2,m_b^2,m_t^2)] \nonumber\\
\frac{\partial \Sigma^{ww}_T(k^2)}{\partial k^2} &=&
 - \frac{g^2 N_c}{96\pi^2 k^2} [-(m_t^2 - m_b^2) (B_0 + 2 B_1) +
 ((m_t^2 - m_b^2)^2 + k^2 (m_t^2 + m_b^2)) B_0^\prime](k^2, m_b^2, m_t^2)
 \nonumber\\
\Sigma^{ww}_L(k^2) &=& - \frac{g^2 N_c}{48\pi^2 k^2}
 [m_t^2 + m_b^2 - 2 A_0(m_t^2) + 2 m_b^2 B_0(k^2,m_b^2,m_t^2) -
 4 (m_t^2 - m_b^2) B_1(k^2,m_b^2,m_t^2)] \nonumber\\
\Sigma^{v_a v_b}_T(k^2) &=& - \frac{N_c}{24\pi^2}
 [2 (G^{L v_a}_{\bar{q} q} G^{L v_b}_{\bar{q} q} +
 G^{R v_a}_{\bar{q} q} G^{R v_b}_{\bar{q} q}) (m_q^2 - A_0(m_q^2)) \nonumber\\
&&- (G^{L v_a}_{\bar{q} q}(G^{L v_b}_{\bar{q} q} - 3 G^{R v_b}_{\bar{q} q}) +
 G^{R v_a}_{\bar{q} q}(G^{R v_b}_{\bar{q} q} - 3 G^{L v_b}_{\bar{q} q}))
 m_q^2 B_0(k^2,m_q^2,m_q^2)] \nonumber\\
\frac{\partial \Sigma^{v_a v_b}_T(k^2)}{\partial k^2} &=&
 -\frac{N_c}{24\pi^2 k^2}
 [(G^{L v_a}_{\bar{q} q} G^{L v_b}_{\bar{q} q} +
 G^{R v_a}_{\bar{q} q} G^{R v_b}_{\bar{q} q}) (- A_0(m_q^2) +
 m_q^2 (1 + B_0(k^2,m_q^2,m_q^2))) \nonumber\\
&&+ 3 (G^{L v_a}_{\bar{q} q} G^{R v_b}_{\bar{q} q} +
 G^{R v_a}_{\bar{q} q} G^{L v_b}_{\bar{q} q})
 m_q^2 k^2 B_0^\prime(k^2,m_q^2,m_q^2)] \nonumber\\
\Sigma^{v_a v_b}_L(k^2) &=& \frac{N_c}{12\pi^2 k^2} [(G^{L v_a}_{\bar{q} q}
 G^{L v_b}_{\bar{q} q} +  G^{R v_a}_{\bar{q} q} G^{R v_b}_{\bar{q} q})
 (-A_0(m_q^2) + m_q^2 (1 + B_0(k^2,m_q^2,m_q^2)))]
\end{eqnarray}
with $q = t,b$.

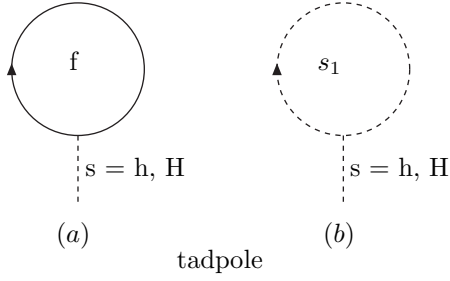
\begin{figure}
\begin{center}
\begin{picture}(200,100)(0,0)
\DashLine(50,15)(50,40){2}
\ArrowArcn(50,65)(25,360,0)
\Text(50,65)[br]{f}
\Text(55,0)[br]{$(a)$}
\Text(53,25)[bl]{s = h, H}
\DashLine(150,15)(150,40){2}
\DashArrowArcn(150,65)(25,360,0){2}
\Text(150,65)[br]{$s_1$}
\Text(155,0)[br]{$(b)$}
\Text(153,25)[bl]{s = h, H}
\Text(120,-10)[br]{tadpole}
\end{picture}
\end{center}
\caption[]{Tadpole diagrams with $s = h, H$.}
\label{tadpole}
\end{figure}
In order to  renormalize the Higgs sector,
we need to calculate $T_s$ ($s=h,H$), which is the summation of
the contribution of two tadpole diagrams
in {\bf Fig}.~\ref{tadpole} ($a$) and ($b$),
$T_s = T_s^{(a)} + T_s^{(b)}$,
where $T_s^{(a)}$, $T_s^{(b)}$ are expressed as
\begin{eqnarray}
T_s^{(a)} &=& - i \frac{N_c}{8\pi^2} (G^{Ls}_{\bar{f}f} + G^{Rs}_{\bar{f}f})
 m_f A_0(m_f^2) \nonumber\\
T_s^{(b)} &=& i \frac{N_c}{16\pi^2} G^{s}_{s_1 s_1} A_0(m_{s_1}^2)
\end{eqnarray}
with $f=t, b$ and $s_1 = \tilde{t}_i, \tilde{b}_i$ ($i=1,2$).

\begin{figure}
\begin{center}
\begin{picture}(360,100)(0,0)
\DashLine(0,50)(25,50){2}
\Text(0,40)[bl]{$s_a$}
\ArrowLine(5,55)(20,55)
\Text(10,60)[bl]{$k$}
\ArrowArcn(50,50)(25,0,180)
\Text(53,62)[br]{$f_1$}
\ArrowArcn(50,50)(25,180,360)
\Text(53,39)[tr]{$f_2$}
\DashLine(75,50)(100,50){2}
\Text(100,40)[br]{$s_b$}
\Text(58,15)[tr]{$(a)$}
\DashLine(120,50)(145,50){2}
\Text(120,40)[bl]{$s_a$}
\DashArrowArcn(170,50)(25,0,180){2}
\Text(173,62)[br]{$s_1$}
\DashArrowArcn(170,50)(25,180,360){2}
\Text(173,39)[tr]{$s_2$}
\DashLine(195,50)(220,50){2}
\Text(220,40)[br]{$s_b$}
\Text(178,20)[tr]{$(b)$}
\DashLine(240,30)(320,30){2}
\Text(240,20)[bl]{$s_a$}
\Text(320,20)[br]{$s_b$}
\DashArrowArcn(280,55)(25,360,0){2}
\Text(270,55)[tr]{$s$}
\Text(285,15)[tr]{$(c)$}
\Text(190,-5)[br]{${\rm s - s}$}
\end{picture}
\end{center}
\caption[]{Higgs boson self-energy ($s_as_b$).}
\label{ss}
\end{figure}
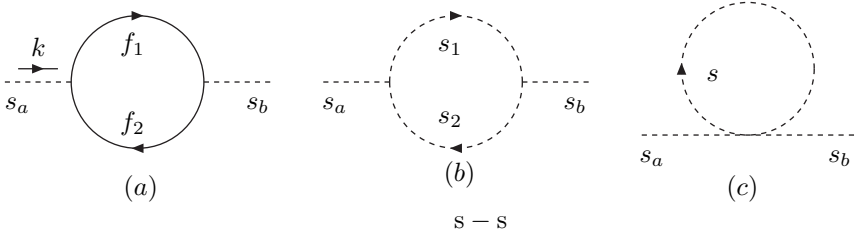
The unrenormalized Higgs boson$-$Higgs boson
$s_a s_b$ ($s_a s_b = hh, hH, HH, GG, GA, AA$)
self-energies are the summation of contribution
shown in {\bf Fig}.~\ref{ss} $a$, $b$ and $c$.
\begin{eqnarray}
\Sigma^{\rm s_a s_b} &=& i \Sigma^{s_a s_b} (k^2) =
 i ( \Sigma^{\rm s_a s_b}_{(a)} (k^2) +
 \Sigma^{\rm s_a s_b}_{(b)} (k^2) + \Sigma^{\rm s_a s_b}_{(c)} (k^2) )
\end{eqnarray}
with
\begin{eqnarray}
\Sigma^{\rm s_a s_b}_{(a)} (k^2) &=& \frac{N_c}{48\pi^2}
 [6 (G^{Ls_a}_{\bar{f}_1 f_2}
 G^{Ls_b}_{\bar{f}_2 f_1} + G^{Rs_a}_{\bar{f}_1 f_2} G^{Rs_b}_{\bar{f}_2 f_1})
 m_{f_1} m_{f_2} B_0 \nonumber\\
&& + (G^{Rs_a}_{\bar{f}_1 f_2} G^{Ls_b}_{\bar{f}_2 f_1} +
 G^{Ls_a}_{\bar{f}_1 f_2} G^{Rs_b}_{\bar{f}_2 f_1}) (24 B_{00} +
 + 6 (B_1 + B_{11}) k^2)] (k^2,m_{f_1}^2,m_{f_2}^2) \nonumber\\
\Sigma^{\rm s_a s_b}_{(b)} (k^2) &=& - \frac{N_c}{16\pi^2} G^{s_a}_{s_1 s_2}
 G^{s_b}_{s_2 s_1} B_0(k^2, m_{s_1}^2, m_{s_2}^2) \nonumber\\
\Sigma^{\rm s_a s_b}_{(c)} (k^2) &=& i \frac{N_c}{16\pi^2}
 G^{s_a s_b}_{s s} A_0(m_s^2)
\end{eqnarray}
where $f_1 f_2 = tt, bb$,
$s_1 s_2 = \tilde{t}_i \tilde{t}_j, \tilde{b}_i \tilde{b}_j$,
and $s = \tilde{t}_i, \tilde{b}_i$ ($i,j=1,2$).

\begin{figure}
\begin{center}
\begin{picture}(100,100)(0,0)
\Photon(0,50)(25,50){2}{5}
\Text(0,40)[bl]{$v$}
\ArrowLine(5,55)(20,55)
\Text(10,60)[bl]{$k$}
\ArrowArcn(50,50)(25,0,180)
\Text(53,62)[br]{$f_1$}
\ArrowArcn(50,50)(25,180,360)
\Text(53,39)[tr]{$f_2$}
\DashLine(75,50)(100,50){2}
\Text(100,40)[br]{$s$}
\Text(34,10)[bl]{${\rm v - s}$}
\end{picture}
\end{center}
\caption[]{Gauge boson$-$Higgs boson mixing ($vs$).}
\label{vs}
\end{figure}
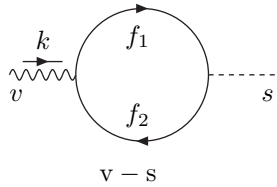
The unrenormalized gauge boson$-$Higgs boson mixing ($vs = zG, zA$)
in {\bf Fig}.~\ref{vs} are expressed as
\begin{eqnarray}
\Sigma^{\rm vs} &=& i k_\mu \Sigma^{\rm vs} (k^2)
\end{eqnarray}
with
\begin{eqnarray}
\Sigma^{\rm vs} (k^2) &=& - \frac{N_c}{8\pi^2} [
 (G^{L v}_{\bar{f}_1 f_2} G^{R s}_{\bar{f}_2 f_1}
 + G^{R v}_{\bar{f}_1 f_2} G^{L s}_{\bar{f}_2 f_1})
 m_{f_1} (B_0 + B_1) \nonumber\\
&& + (G^{L v}_{\bar{f}_1 f_2} G^{L s}_{\bar{f}_2 f_1} +
 G^{R v}_{\bar{f}_1 f_2} G^{R s}_{\bar{f}_2 f_1}) m_{f_2} B_1 ]
 (k^2,m_{f_1}^2,m_{f_2}^2)
\end{eqnarray}
where $f_1 f_2 = tt, bb$.

\begin{figure}
\begin{center}
\begin{picture}(100,100)(0,0)
\ArrowLine(5,30)(25,30)
\Text(5,20)[bl]{$t$}
\ArrowLine(5,35)(20,35)
\Text(10,40)[bl]{$k$}
\ArrowLine(25,30)(75,30)
\Text(45,18)[bl]{$f$}
\ArrowLine(75,30)(95,30)
\Text(75,20)[br]{$t$}
\DashArrowArcn(50,30)(25,-180,0){2}
\Text(34,0)[bl]{${\rm t - t}$}
\end{picture}
\end{center}
\caption[]{Top quark self-energy.}
\label{tt}
\end{figure}
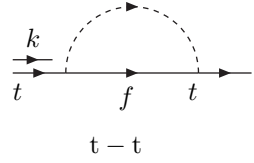
The unrenormalized top quark self-energy shown in {\bf Fig}.~\ref{tt}
are expressed as
\begin{eqnarray}
\Sigma^{\rm tt} &=& i[\Sigma^t_L(k^2) \dslash{k} P_L +
 \Sigma^t_R(k^2) \dslash{k} P_R + \Sigma^t_S(k^2)] \nonumber\\
\end{eqnarray}
with
\begin{eqnarray}
\Sigma^t_S(k^2) &=& -\frac{1}{16\pi^2} G^{Ls}_{\bar{t}f} G^{Ls}_{\bar{f}t}
 m_f B_0(k^2,m_f^2,m_s^2) \nonumber\\
\Sigma^t_L(k^2) &=& \frac{1}{16\pi^2} G^{Rs}_{\bar{t}f} G^{Ls}_{\bar{f}t}
 B_1(k^2,m_f^2,m_s^2) \nonumber\\
\Sigma^t_R(k^2) &=& \frac{1}{16\pi^2} G^{Ls}_{\bar{t}f} G^{Rs}_{\bar{f}t}
 B_1(k^2,m_f^2,m_s^2)
\end{eqnarray}
where $f s = th, tH, tG, tA, bG^+, bH^+,
\tilde{\chi}_i^+ \tilde{b}_\alpha, \tilde{\chi}_j^0 \tilde{t}_\alpha$,
($i, \alpha = 1, 2$, $j=1,2,3,4$).

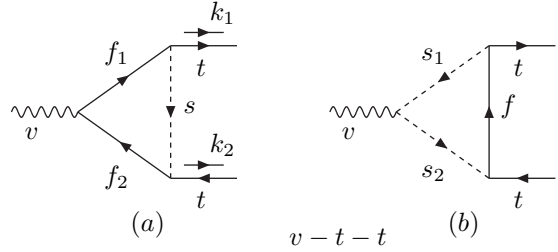
\begin{figure}
\begin{center}
\begin{picture}(240,100)(0,0)
\Photon(0,50)(25,50){2}{5}
\Text(5,40)[bl]{$v$}
\ArrowLine(25,50)(60,75)
\Text(35,70)[bl]{$f_1$}
\ArrowLine(60,25)(25,50)
\Text(35,30)[tl]{$f_2$}
\DashArrowLine(60,75)(60,25){2}
\Text(65,50)[bl]{$s$}
\ArrowLine(60,75)(85,75)
\Text(70,70)[tl]{$t$}
\ArrowLine(85,25)(60,25)
\Text(70,20)[tl]{$t$}
\Text(45,5)[bl]{$(a)$}
\ArrowLine(65,80)(80,80)
\Text(75,85)[bl]{$k_1$}
\ArrowLine(65,30)(80,30)
\Text(75,35)[bl]{$k_2$}
\Photon(120,50)(145,50){2}{5}
\Text(125,40)[bl]{$v$}
\DashArrowLine(180,75)(145,50){2}
\Text(155,70)[bl]{$s_1$}
\DashArrowLine(145,50)(180,25){2}
\Text(155,30)[tl]{$s_2$}
\ArrowLine(180,25)(180,75)
\Text(185,50)[bl]{$f$}
\ArrowLine(180,75)(205,75)
\Text(190,70)[tl]{$t$}
\ArrowLine(205,25)(180,25)
\Text(190,20)[tl]{$t$}
\Text(165,5)[bl]{$(b)$}
\Text(105,0)[bl]{$v-t-t$}
\end{picture}
\end{center}
\caption[]{Gauge boson$-$top quark$-$top quark vertex ($vtt$).}
\label{vtt}
\end{figure}
The unrenormalized gauge boson$-$top quark$-$top quark vertex
($vtt = \gamma t t$, $Ztt$)
shown in {\bf Fig}.~\ref{vtt} ($a$) and ($b$) are
\begin{eqnarray}
\Gamma^{\rm vtt} &=& \Gamma^L_1 {k_1}_\mu P_L +
 \Gamma^L_2 {k_2}_\mu P_L + \Gamma^L_3 \gamma_\mu P_L +
 \Gamma^L_4 \dslash{k_1} {k_1}_\mu P_L +
 \Gamma^L_5 \dslash{k_1} {k_2}_\mu P_L +
 \Gamma^L_6 \dslash{k_2} {k_1}_\mu P_L + \nonumber\\
&& \Gamma_7 \dslash{k_2} {k_2}_\mu P_L +
 \Gamma^L_8 \dslash{k_1} \gamma_\mu P_L
 + \Gamma^L_9 \dslash{k_2} \gamma_\mu P_L
 + \Gamma^L_{10} \dslash{k_1} \dslash{k_2} \gamma_\mu P_L + (L \rightarrow R)
\end{eqnarray}
where $\Gamma^{L,R}_i$ ($i=1,2, \dots, 10$)
is the summation of $\Gamma^{L,R}_{i(a)}$
and $\Gamma^{L,R}_{i(b)}$,
$\Gamma^{L,R}_i = \Gamma^{L,R}_{i(a)} + \Gamma^{L,R}_{i(b)}$,
where $\Gamma^L_{i(a)}$ and $\Gamma^L_{i(b)}$ are expressed as follow,
and we get $\Gamma^R_{i(a,b)}$ through the exchange of ($L\leftrightarrow R$)
in the expressions of corresponding $\Gamma^L_{i(a,b)}$.
\begin{eqnarray}
\Gamma^L_{1(a)} &=&  \frac{1}{8\pi^2} G^{R v}_{\bar{f}_1 f_2}
 G^{L s}_{\bar{t} f_1} G^{L s}_{\bar{f}_2 t} m_{f_1} C_2 \nonumber\\
\Gamma^L_{2(a)} &=& -
 \frac{1}{8\pi^2} G^{R v}_{\bar{f}_1 f_2} G^{L s}_{\bar{t} f_1}
 G^{L s}_{\bar{f}_2 t} m_{f_1} (C_0 + C_1) \nonumber\\
\Gamma^L_{3(a)} &=& \frac{1}{32\pi^2} G^{R s}_{\bar{t} f_1}
 G^{L s}_{\bar{f}_2 t} (2 G^{L v}_{\bar{f}_1 f_2} m_{f_1} m_{f_2} C_0 -
 G^{R v}_{\bar{f}_1 f_2} (4 C_{00} + 2 (C_2 + C_{22}) k_1^2 - \nonumber\\
&& 4 C_{12} k_1 \cdot k_2 + 2 (C_1 + C_{11}) k_2^2)) \nonumber\\
\Gamma^L_{4(a)} &=&  \frac{1}{8\pi^2} G^{R v}_{\bar{f}_1 f_2}
 G^{R s}_{\bar{t} f_1} G^{L s}_{\bar{f}_2 t} (C_2 + C_{22}) \nonumber\\
\Gamma^L_{5(a)} &=& - \frac{1}{8\pi^2} G^{R v}_{\bar{f}_1 f_2}
 G^{R s}_{\bar{t} f_1}
 G^{L s}_{\bar{f}_2 t} (C_0 + C_1 + C_{12} + C_2) \nonumber\\
\Gamma^L_{6(a)} &=&  - \frac{1}{8\pi^2}
 G^{R v}_{\bar{f}_1 f_2} G^{R s}_{\bar{t} f_1}
 G^{L s}_{\bar{f}_2 t} C_{12} \nonumber\\
\Gamma^L_{7(a)} &=& \frac{1}{8\pi^2} G^{R v}_{\bar{f}_1 f_2}
 G^{R s}_{\bar{t} f_1} G^{L s}_{\bar{f}_2 t} (C_1 + C_{11}) \nonumber\\
\Gamma^L_{8(a)} &=& \frac{1}{16\pi^2} G^{L s}_{\bar{t} f_1}
 G^{L s}_{\bar{f}_2 t} (G^{L v}_{\bar{f}_1 f_2} m_{f_2} C_0 -
 G^{R v}_{\bar{f}_1 f_2} m_{f_1} C_2 +
 G^{L v}_{\bar{f}_1 f_2} m_{f_2} C_2) \nonumber\\
\Gamma^L_{9(a)} &=& \frac{1}{16\pi^2} G^{L s}_{\bar{t} f_1}
 G^{L s}_{\bar{f}_2 t} (G^{R v}_{\bar{f}_1 f_2} m_{f_1} C_0 +
 G^{R v}_{\bar{f}_1 f_2} m_{f_1} C_1 -
 G^{L v}_{\bar{f}_1 f_2} m_{f_2} C_1) \nonumber\\
\Gamma^L_{10(a)} &=& \frac{1}{16\pi^2} G^{R v}_{\bar{f}_1 f_2}
 G^{R s}_{\bar{t} f_1} G^{L s}_{\bar{f}_2 t} (C_0 + C_1 + C_2) \nonumber\\
\Gamma^L_{1(b)} &=&  -\frac{1}{16\pi^2} G^{v}_{s_2 s_1} G^{L s_1}_{\bar{t} f}
 G^{L s_2}_{\bar{f} t} m_f (C_0 + 2 C_2) \nonumber\\
\Gamma^L_{2(b)} &=& \frac{1}{16\pi^2} G^{v}_{s_2 s_1} G^{L s_1}_{\bar{t} f}
 G^{L s_2}_{\bar{f} t} m_f (C_0 + 2 C_1) \nonumber\\
\Gamma^L_{3(b)} &=& \frac{1}{8\pi^2} G^{v}_{s_2 s_1} G^{R s_1}_{\bar{t} f}
 G^{L s_2}_{\bar{f} t} C_{00} \nonumber\\
\Gamma^L_{4(b)} &=&   \frac{1}{16\pi^2} G^{v}_{s_2 s_1} G^{R s_1}_{\bar{t} f}
 G^{L s_2}_{\bar{f} t} (C_2 + 2 C_{22}) \nonumber\\
\Gamma^L_{5(b)} &=& - \frac{1}{16\pi^2} G^{v}_{s_2 s_1} G^{R s_1}_{\bar{t} f}
 G^{L s_2}_{\bar{f} t} (2 C_{12} + C_2) \nonumber\\
\Gamma^L_{6(b)} &=&  -\frac{1}{16\pi^2} G^{v}_{s_2 s_1} G^{R s_1}_{\bar{t} f}
 G^{L s_2}_{\bar{f} t} {k_1}_\mu (C_1 + 2C_{12}) \nonumber\\
\Gamma^L_{7(b)} &=& \frac{1}{16\pi^2} G^{v}_{s_2 s_1} G^{R s_1}_{\bar{t} f}
 G^{L s_2}_{\bar{f} t} (C_1 + 2 C_{11}) \nonumber\\
\Gamma^L_{8(b)} &=&  0 \nonumber\\
\Gamma^L_{9(b)} &=&  0 \nonumber\\
\Gamma^L_{10(b)} &=&  0
\end{eqnarray}
with the auguments of the $C$ function as
$C(k_2^2, (k_1+k_2)^2, k_1^2, m_s^2, m_{f_2}^2, m_{f_1}^2)$
and $C(k_2^2, (k_1+k_2)^2, k_1^2, m_f^2, m_{s_2}^2, m_{s_1}^2)$
for {\bf Fig}.~\ref{vtt} ($a$) and ($b$) respectively.
For both vertex $\gamma tt$ and $Z tt$,
the virtual particles propogated in the loops are as follow,
$f_1 f_2 s = tth, ttH, ttG, ttA, bbG^-, bbH^-$,
$f s_1 s_2 = bG^-G^-, bH^-H^-$,
and additionally $f_1 f_2 s =
\tilde{\chi}^+_i \tilde{\chi}^+_i \tilde{b}^\ast_\alpha$,
$f s_1 s_2 =
\tilde{\chi}^+_i \tilde{b}^\ast_\alpha \tilde{b}^\ast_\alpha,
\tilde{\chi}^0_k \tilde{t}^\ast_\alpha \tilde{t}^\ast_\alpha$
for $\gamma tt$ vertex,
$f_1 f_2 s =
\tilde{\chi}^+_i \tilde{\chi}^+_j \tilde{b}^\ast_\alpha,
\tilde{\chi}^0_k \tilde{\chi}^0_l \tilde{t}^\ast_\alpha$,
$f s_1 s_2 = thA, tAh, tHA, tAH, thG, tGh, tHG,\\ tGH,
\tilde{\chi}^+_i \tilde{b}^\ast_\alpha \tilde{b}^\ast_\beta,
\tilde{\chi}^0_k \tilde{t}^\ast_\alpha \tilde{t}^\ast_\beta$
for $Z tt$ vertex, with $i,j=1,2$, $\alpha,\beta=1,2$, $k,l=1,2,3,4$.

\begin{figure}
\begin{center}
\begin{picture}(240,100)(0,0)
\DashArrowLine(0,50)(25,50){2}
\Text(5,40)[bl]{$s$}
\ArrowLine(25,50)(60,75)
\Text(35,70)[bl]{$f_1$}
\ArrowLine(60,25)(25,50)
\Text(35,30)[tl]{$f_2$}
\DashArrowLine(60,75)(60,25){2}
\Text(65,50)[bl]{$s_1$}
\ArrowLine(60,75)(85,75)
\Text(70,70)[tl]{$t$}
\ArrowLine(85,25)(60,25)
\Text(70,20)[tl]{$t$}
\Text(55,5)[bl]{$(a)$}
\ArrowLine(65,80)(80,80)
\Text(75,85)[bl]{$k_1$}
\ArrowLine(65,30)(80,30)
\Text(75,35)[bl]{$k_2$}
\DashArrowLine(120,50)(145,50){2}
\Text(125,40)[bl]{$s$}
\DashArrowLine(180,75)(145,50){2}
\Text(155,70)[bl]{$s_1$}
\DashArrowLine(145,50)(180,25){2}
\Text(155,30)[tl]{$s_2$}
\ArrowLine(180,25)(180,75)
\Text(185,50)[bl]{$f$}
\ArrowLine(180,75)(205,75)
\Text(190,70)[tl]{$t$}
\ArrowLine(205,25)(180,25)
\Text(190,20)[tl]{$t$}
\Text(175,5)[bl]{$(b)$}
\Text(105,0)[bl]{$s-t-t$}
\end{picture}
\end{center}
\caption[]{Higgs boson$-$top quark$-$top quark vertex ($stt$).}
\label{stt}
\end{figure}
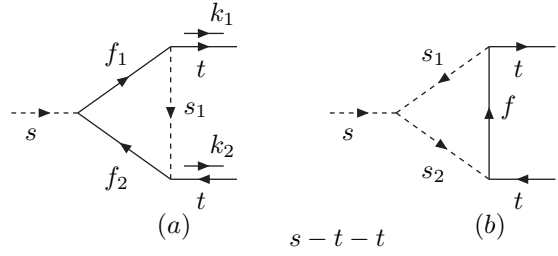
The unrenormalized Higgs boson$-$top quark$-$top quark
($stt = htt, Gtt, Att$) vertex shown
in {\bf Fig}.~\ref{stt} ($a$) and ($b$) are
\begin{eqnarray}
\Gamma^{\rm stt} &=& \Gamma^L_1 P_L + \Gamma^L_2 \dslash{k_1} P_L +
 \Gamma^L_3\dslash{k_2} P_L
 + \Gamma^L_4 \dslash{k_1} \dslash{k_2} P_L + (L \rightarrow R)
\end{eqnarray}
where $\Gamma^{L,R}_i$ ($i=1,2,3,4$)
is the summation of $\Gamma^{L,R}_{i(a)}$
and $\Gamma^{L,R}_{i(b)}$,
$\Gamma^{L,R}_i = \Gamma^{L,R}_{i(a)} + \Gamma^{L,R}_{i(b)}$,
where $\Gamma^L_{i(a)}$ and $\Gamma^L_{i(b)}$ are expressed as follow,
and we get $\Gamma^R_{i(a,b)}$ through the exchange of ($L\leftrightarrow R$)
in the expressions of corresponding $\Gamma^L_{i(a,b)}$.
\begin{eqnarray}
\Gamma^L_{1(a)} &=& \frac{1}{32\pi^2} G^{L s_1}_{\bar{t} f_1}
 G^{L s_1}_{\bar{f}_2 t} (2 G^{L s}_{\bar{f}_1 f_2} m_{f_1} m_{f_2} C_0 +
 G^{R s}_{\bar{f}_1 f_2} (8 C_{00} + 2 (C_2 + C_{22}) k_1^2 - \nonumber\\
&& 4 C_{12} k_1 \cdot k_2 + 2 (C_1 + C_{11}) k_2^2)) \nonumber\\
\Gamma^L_{2(a)} &=& \frac{1}{16\pi^2} G^{R s_1}_{\bar{t} f_1}
 G^{L s_1}_{\bar{f}_2 t} (G^{L s}_{\bar{f}_1 f_2} m_{f_2} C_0 +
 G^{R s}_{\bar{f}_1 f_2} m_{f_1} C_2 +
 G^{L s}_{\bar{f}_1 f_2} m_{f_2} C_2) \nonumber\\
\Gamma^L_{3(a)} &=& - \frac{1}{16\pi^2} G^{R s_1}_{\bar{t} f_1}
 G^{L s_1}_{\bar{f}_2 t} (G^{R s}_{\bar{f}_1 f_2} m_{f_1} C_0 +
 G^{R s}_{\bar{f}_1 f_2} m_{f_1} C_1 +
 G^{L s}_{\bar{f}_1 f_2} m_{f_2} C_1) \nonumber\\
\Gamma^L_{4(a)} &=&  -\frac{1}{16\pi^2} G^{R s}_{\bar{f}_1 f_2}
 G^{L s_1}_{\bar{t} f_1} G^{L s_1}_{\bar{f}_2 t} (C_0 + C_1 + C_2) \nonumber\\
\Gamma^L_{1(b)} &=& \frac{1}{16\pi^2} G^{s}_{s_2 s_1} G^{L s_1}_{\bar{t} f}
 G^{L s_2}_{\bar{f} t} m_f C_0 \nonumber\\
\Gamma^L_{2(b)} &=& -\frac{1}{16\pi^2} G^{s}_{s_2 s_1} G^{R s_1}_{\bar{t} f}
 G^{L s_2}_{\bar{f} t} C_2 \nonumber\\
\Gamma^L_{3(b)} &=& \frac{1}{16\pi^2} G^{s}_{s_2 s_1} G^{R s_1}_{\bar{t} f}
 G^{L s_2}_{\bar{f} t} C_1 \nonumber\\
\Gamma^L_{4(b)} &=& 0
\end{eqnarray}
with the auguments of the $C$ function as
$C(k_2^2, (k_1+k_2)^2, k_1^2, m_s^2, m_{f_2}^2, m_{f_1}^2)$
and $C(k_2^2, (k_1+k_2)^2, k_1^2, m_f^2, m_{s_2}^2, m_{s_1}^2)$
for {\bf Fig}.~\ref{stt} ($a$) and ($b$) respectively.
For vertex $htt$, $Gtt$ and $Att$,
the virtual particles propogated in the loops are as follow,
$f_1 f_2 s = tth, ttH, ttG, ttA, bbG^-, bbH^-,
\tilde{\chi}^+_i \tilde{\chi}^+_j \tilde{b}^\ast_\alpha,
\tilde{\chi}^0_k \tilde{\chi}^0_l \tilde{t}^\ast_\alpha$,
$f s_1 s_2 =
\tilde{\chi}^+_i \tilde{b}^\ast_\alpha \tilde{b}^\ast_\alpha,
\tilde{\chi}^0_k \tilde{t}^\ast_\alpha \tilde{t}^\ast_\alpha$,
and additionally $f s_1 s_2 =
thh, thH, tHh, tHH, tGG, tGA, tAG, tAA, bG^-G^-, bG^-H^-, bH^-G^-, bH^-H^-$
for $htt$ vertex,
$f s_1 s_2 = thA, tAh, tHA, tAH, thG, tGh, tHG, tGH$
for $G tt$ vertex,
and $f s_1 s_2 = thA, tAh, tHA, tAH, thG, tGh, tHG, tGH,\\ bG^-H^-, bH^-G^-$
for $A tt$ vertex, with $i,j=1,2$, $\alpha,\beta=1,2$, $k,l=1,2,3,4$,

\begin{figure}
\begin{center}
\begin{picture}(360,200)(0,0)
\DashArrowLine(0,150)(25,150){2}
\Text(5,140)[bl]{$s$}
\ArrowLine(25,150)(60,175)
\Text(35,170)[bl]{$f_1$}
\ArrowLine(60,125)(25,150)
\Text(35,130)[tl]{$f_2$}
\ArrowLine(60,175)(60,125)
\Text(65,150)[bl]{$f_3$}
\Photon(60,175)(85,175){2}{5}
\Text(70,170)[tl]{$v_a$}
\Photon(85,125)(60,125){2}{5}
\Text(70,120)[tl]{$v_b$}
\Text(55,105)[bl]{$(a)$}
\ArrowLine(65,180)(80,180)
\Text(75,185)[bl]{$k_1$}
\ArrowLine(65,130)(80,130)
\Text(75,135)[bl]{$k_2$}
\DashArrowLine(120,150)(145,150){2}
\Text(125,140)[bl]{$s$}
\ArrowLine(180,175)(145,150)
\Text(155,170)[bl]{$f_1$}
\ArrowLine(145,150)(180,125)
\Text(155,130)[tl]{$f_2$}
\ArrowLine(180,125)(180,175)
\Text(185,150)[bl]{$f_3$}
\Photon(180,175)(205,175){2}{5}
\Text(190,170)[tl]{$v_a$}
\Photon(205,125)(180,125){2}{5}
\Text(190,120)[tl]{$v_b$}
\Text(175,105)[bl]{$(b)$}
\DashArrowLine(0,50)(25,50){2}
\Text(5,40)[bl]{$s$}
\DashArrowLine(25,50)(60,75){2}
\Text(35,70)[bl]{$s_1$}
\DashArrowLine(60,25)(25,50){2}
\Text(35,30)[tl]{$s_2$}
\DashArrowLine(60,75)(60,25){2}
\Text(65,50)[bl]{$s_3$}
\Photon(60,75)(85,75){2}{5}
\Text(70,70)[tl]{$v_a$}
\Photon(85,25)(60,25){2}{5}
\Text(70,20)[tl]{$v_b$}
\Text(55,5)[bl]{$(c)$}
\DashArrowLine(120,50)(145,50){2}
\Text(125,40)[bl]{$s$}
\DashArrowLine(180,75)(145,50){2}
\Text(155,70)[bl]{$s_1$}
\DashArrowLine(145,50)(180,25){2}
\Text(155,30)[tl]{$s_2$}
\DashArrowLine(180,25)(180,75){2}
\Text(185,50)[bl]{$s_3$}
\Photon(180,75)(205,75){2}{5}
\Text(190,70)[tl]{$v_a$}
\Photon(205,25)(180,25){2}{5}
\Text(190,20)[tl]{$v_b$}
\Text(170,10)[bl]{$(d)$}
\DashArrowLine(240,100)(265,100){2}
\DashArrowArcn(290,100)(25,0,180){2}
\Text(293,112)[br]{$s_1$}
\DashArrowArcn(290,100)(25,180,360){2}
\Text(293,89)[tr]{$s_2$}
\Photon(315,100)(340,125){2}{5}
\Text(335,110)[bl]{$v_a$}
\Photon(315,100)(340,75){2}{5}
\Text(335,90)[tl]{$v_b$}
\Text(285,55)[bl]{$(e)$}
\Text(165,-5)[bl]{$s-v-v$}
\end{picture}
\end{center}
\caption[]{Higgs boson$-$gauge boson$-$gauge boson vertex ($sv_av_b$).}
\label{svv}
\end{figure}
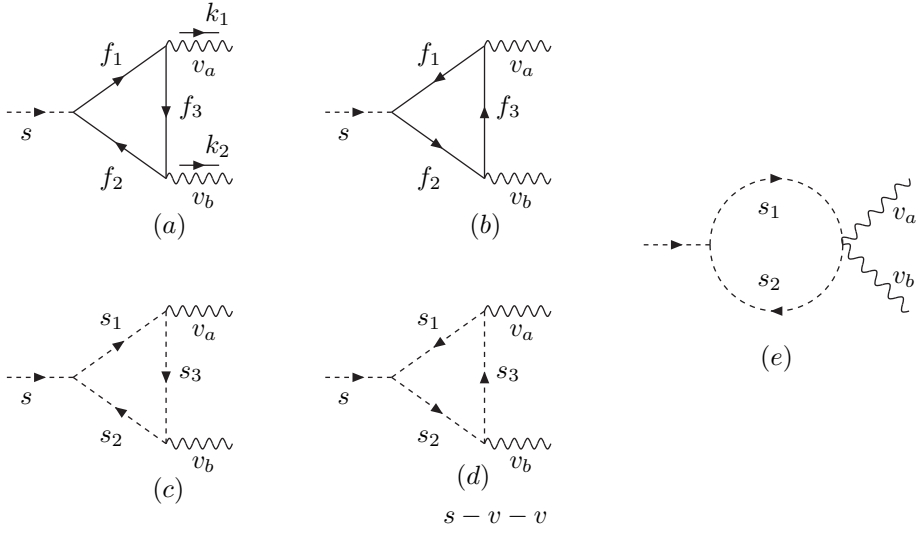
The unrenormalized Higgs boson$-$gauge boson$-$gauge boson
($s v_a v_b = h \gamma \gamma, h \gamma Z, h Z Z$)
vertex shown in {\bf Fig}.~\ref{svv} ($a$)-($e$) are
\begin{eqnarray}
\Gamma^{\rm s v_a v_b} &=& \Gamma_1 g_{\mu\nu} +
 \Gamma_2 {k_1}_\mu {k_1}_\nu + \Gamma_3 {k_2}_\mu {k_2}_\nu
 + \Gamma_4 {k_1}_\mu {k_2}_\nu + \Gamma_5 {k_2}_\mu {k_1}_\nu
 + \Gamma_6 \epsilon^{\mu\nu\sigma\tau} {k_1}_\sigma {k_2}_\tau
\end{eqnarray}
where $\Gamma_i$ ($i=1,2,\dots,6$)
is the summation from $\Gamma_{i(a)}$ to $\Gamma_{i(e)}$,
$\Gamma_i = \Gamma_{i(a)} + \Gamma_{i(b)} + \dots + \Gamma_{i(e)}$,
where $\Gamma_{i(a)}$, $\Gamma_{i(c)}$ and
$\Gamma_{i(e)}$ are expressed as follow,
\begin{eqnarray}
\Gamma_{1(a)} &=& \frac{1}{16\pi^2} [-2 (G^{L s}_{\bar{f}_1 f_2}
 G^{L v_a}_{\bar{f}_3 f_1} G^{R v_b}_{\bar{f}_2 f_3} +
 G^{R s}_{\bar{f}_1 f_2} G^{R v_a}_{\bar{f}_3 f_1}
 G^{L v_b}_{\bar{f}_2 f_3}) m_{f_1} m_{f_2} m_{f_3} C_0 -
 (G^{R s}_{\bar{f}_1 f_2} G^{L v_a}_{\bar{f}_3 f_1}
 G^{L v_b}_{\bar{f}_2 f_3} + G^{L s}_{\bar{f}_1 f_2}
 G^{R v_a}_{\bar{f}_3 f_1} G^{R v_b}_{\bar{f}_2 f_3}) m_{f_2} \nonumber\\
&&  (- 4 C_{00} - 2 (C_2 + C_{22}) k_1^2 + 2 (C_1 + 2 C_{12}) k_1 \cdot k_2 -
 2 C_{11} k_2^2) + \nonumber\\
&& (G^{R s}_{\bar{f}_1 f_2} G^{L v_a}_{\bar{f}_3 f_1}
 G^{R v_b}_{\bar{f}_2 f_3} + G^{L s}_{\bar{f}_1 f_2}
 G^{R v_a}_{\bar{f}_3 f_1} G^{L v_b}_{\bar{f}_2 f_3})
 m_{f_3} (- 8 C_{00} - 2 (C_2 + C_{22}) k_1^2 +
\nonumber\\
&&  2 (C_0 + C_1 + 2 C_{12} + C_2) k_1 \cdot k_2 - 2 (C_1 + C_{11}) k_2^2) +
 (G^{L s}_{\bar{f}_1 f_2} G^{L v_a}_{\bar{f}_3 f_1}
 G^{L v_b}_{\bar{f}_2 f_3} + G^{R s}_{\bar{f}_1 f_2}
 G^{R v_a}_{\bar{f}_3 f_1} G^{R v_b}_{\bar{f}_2 f_3}) m_{f_1} \nonumber\\
&&  (4 C_{00} + 2 C_{22} k_1^2 - 2 (2 C_{12} + C_2) k_1 \cdot k_2 +
 2 (C_1 + C_{11}) k_2^2)] \nonumber\\
\Gamma_{2(a)} &=& - \frac{1}{4\pi^2} [(G^{L s}_{\bar{f}_1 f_2}
 G^{L v_a}_{\bar{f}_3 f_1} G^{L v_b}_{\bar{f}_2 f_3} +
 G^{R s}_{\bar{f}_1 f_2} G^{R v_a}_{\bar{f}_3 f_1}
 G^{R v_b}_{\bar{f}_2 f_3}) m_{f_1} C_{22} +
 (G^{R s}_{\bar{f}_1 f_2} G^{L v_a}_{\bar{f}_3 f_1}
 G^{L v_b}_{\bar{f}_2 f_3} + G^{L s}_{\bar{f}_1 f_2}
 G^{R v_a}_{\bar{f}_3 f_1} G^{R v_b}_{\bar{f}_2 f_3}) m_{f_2} (C_2 + C_{22})]
 \nonumber\\
\Gamma_{3(a)} &=& -\frac{1}{4\pi^2} [ (G^{L s}_{\bar{f}_1 f_2}
 G^{L v_a}_{\bar{f}_3 f_1} G^{L v_b}_{\bar{f}_2 f_3} +
 G^{R s}_{\bar{f}_1 f_2} G^{R v_a}_{\bar{f}_3 f_1}
 G^{R v_b}_{\bar{f}_2 f_3}) m_{f_1} (C_1 + C_{11}) +
 (G^{R s}_{\bar{f}_1 f_2} G^{L v_a}_{\bar{f}_3 f_1}
 G^{L v_b}_{\bar{f}_2 f_3} + G^{L s}_{\bar{f}_1 f_2}
 G^{R v_a}_{\bar{f}_3 f_1} G^{R v_b}_{\bar{f}_2 f_3}) m_{f_2} C_{11}]
 \nonumber\\
\Gamma_{4(a)} &=& \frac{1}{8\pi^2} [ (G^{L s}_{\bar{f}_1 f_2}
 G^{L v_a}_{\bar{f}_3 f_1} G^{L v_b}_{\bar{f}_2 f_3} +
 G^{R s}_{\bar{f}_1 f_2} G^{R v_a}_{\bar{f}_3 f_1}
 G^{R v_b}_{\bar{f}_2 f_3}) m_{f_1} (2 C_{12} + C_2) +
 (G^{R s}_{\bar{f}_1 f_2} G^{L v_a}_{\bar{f}_3 f_1}
 G^{L v_b}_{\bar{f}_2 f_3} + G^{L s}_{\bar{f}_1 f_2}
 G^{R v_a}_{\bar{f}_3 f_1} G^{R v_b}_{\bar{f}_2 f_3}) m_{f_2} \nonumber\\
&& (C_1 + 2 C_{12}) + (G^{R s}_{\bar{f}_1 f_2}
 G^{L v_a}_{\bar{f}_3 f_1} G^{R v_b}_{\bar{f}_2 f_3} +
 G^{L s}_{\bar{f}_1 f_2} G^{L v_b}_{\bar{f}_2 f_3}
 G^{R v_a}_{\bar{f}_3 f_1}) m_{f_3} (C_0 + C_1 + C_2) ] \nonumber\\
\Gamma_{5(a)} &=& \frac{1}{8\pi^2} [ (G^{L s}_{\bar{f}_1 f_2}
 G^{L v_a}_{\bar{f}_3 f_1} G^{L v_b}_{\bar{f}_2 f_3} +
 G^{R s}_{\bar{f}_1 f_2} G^{R v_a}_{\bar{f}_3 f_1}
 G^{R v_b}_{\bar{f}_2 f_3}) m_{f_1} (2 C_{12} + C_2) +
 (G^{R s}_{\bar{f}_1 f_2} G^{L v_a}_{\bar{f}_3 f_1}
 G^{L v_b}_{\bar{f}_2 f_3} + G^{L s}_{\bar{f}_1 f_2}
 G^{R v_a}_{\bar{f}_3 f_1} G^{R v_b}_{\bar{f}_2 f_3}) m_{f_2} \nonumber\\
&& (C_1 + 2 C_{12}) - (G^{R s}_{\bar{f}_1 f_2}
 G^{L v_a}_{\bar{f}_3 f_1} G^{R v_b}_{\bar{f}_2 f_3} +
 G^{L s}_{\bar{f}_1 f_2} G^{R v_a}_{\bar{f}_3 f_1}
 G^{L v_b}_{\bar{f}_2 f_3}) m_{f_3} (C_0 + C_1 + C_2)] \nonumber\\
\Gamma_{6(a)} &=& \frac{i}{8\pi^2} [ (- G^{L s}_{\bar{f}_1 f_2}
 G^{L v_a}_{\bar{f}_3 f_1} G^{L v_b}_{\bar{f}_2 f_3} +
 G^{R s}_{\bar{f}_1 f_2} G^{R v_a}_{\bar{f}_3 f_1}
 G^{R v_b}_{\bar{f}_2 f_3}) m_{f_1} C_2 +
 (G^{R s}_{\bar{f}_1 f_2} G^{L v_a}_{\bar{f}_3 f_1}
 G^{L v_b}_{\bar{f}_2 f_3} - G^{L s}_{\bar{f}_1 f_2}
 G^{R v_a}_{\bar{f}_3 f_1} G^{R v_b}_{\bar{f}_2 f_3})
 m_{f_2} C_1 + \nonumber\\
&& (- G^{R s}_{\bar{f}_1 f_2} G^{L v_a}_{\bar{f}_3 f_1}
 G^{R v_b}_{\bar{f}_2 f_3} + G^{L s}_{\bar{f}_1 f_2}
 G^{R v_a}_{\bar{f}_3 f_1} G^{L v_b}_{\bar{f}_2 f_3}) m_{f_3}
 (C_0 + C_1 + C_2)] \nonumber\\
\Gamma_{1(c)} &=& \frac{1}{4\pi^2} G^{v_a}_{s_3 s_1} G^{v_b}_{s_2 s_3}
 G^{s}_{s_1 s_2} C_{00} \nonumber\\
\Gamma_{2(c)} &=& \frac{1}{8\pi^2} G^{v_a}_{s_3 s_1} G^{v_b}_{s_2 s_3}
 G^{s}_{s_1 s_2} (C_2 + 2 C_{22}) \nonumber\\
\Gamma_{3(c)} &=& \frac{1}{8\pi^2} G^{v_a}_{s_3 s_1} G^{v_b}_{s_2 s_3}
 G^{s}_{s_1 s_2} (C_1 + 2 C_{11}) \nonumber\\
\Gamma_{4(c)} &=& -\frac{1}{16\pi^2} G^{v_a}_{s_3 s_1} G^{v_b}_{s_2 s_3}
 G^{s}_{s_1 s_2} (C_0 + 2 C_1 + 4 C_{12} + 2 C_2) \nonumber\\
\Gamma_{5(c)} &=&  -\frac{1}{4\pi^2} G^{v_a}_{s_3 s_1} G^{v_b}_{s_2 s_3}
 G^{s}_{s_1 s_2} C_{12} \nonumber\\
\Gamma_{6(c)} &=&  0 \nonumber\\
\Gamma_{1(e)} &=& - \frac{1}{16\pi^2} G^{v_a v_b}_{s_2 s_1}
 G^{s}_{s_1 s_2} B_0((k1+k2)^2, m_{s_1}^2, m_{s_2}^2) \nonumber\\
\Gamma_{2(e)} &=& 0 \nonumber\\
\Gamma_{3(e)} &=& 0 \nonumber\\
\Gamma_{4(e)} &=& 0 \nonumber\\
\Gamma_{5(e)} &=& 0 \nonumber\\
\Gamma_{6(e)} &=& 0
\end{eqnarray}
with arguments of $C$ functions as
$C(k_2^2, (k_1+k_2)^2, k_1^2, m_{f_3}^2, m_{f_2}^2, m_{f_1}^2)$
and $C(k_2^2, (k_1+k_2)^2, k_1^2, m_{s_3}^2, m_{s_2}^2, m_{s_1}^2)$
for {\bf Fig}.~\ref{svv} ($a$) and ($c$),
and we get the result of {\bf Fig}.~\ref{svv} ($b$) and ($d$) with
the exchange of $v_a \leftrightarrow v_b$ and $f,s_1 \leftrightarrow f,s_2$
from the one of {\bf Fig}.~\ref{svv} ($a$) and ($c$).
For the vertex $h \gamma \gamma$, $h \gamma Z$, $h Z Z$,
the virtual fields propogated in the loops are as follow,
$f_1 f_2 f_3 = ttt, bbb$ in {\bf Fig}.~\ref{svv} ($a$),
$s_1 s_2 s_3 = \tilde{t}_\alpha \tilde{t}_\beta \tilde{t}_\gamma,
\tilde{b}_\alpha \tilde{b}_\beta \tilde{b}_\gamma$
in {\bf Fig}.~\ref{svv} ($c$),
$s_1 s_2 = \tilde{t}_\alpha \tilde{t}_\beta,
\tilde{b}_\alpha \tilde{b}_\beta$ in {\bf Fig}.~\ref{svv} ($e$).
Note that $\gamma$ couples only with scalar quarks of the same quantum numbers,
i.e., $\gamma \tilde{t}_\alpha \tilde{t}_\alpha$.

\begin{figure}
\begin{center}
\begin{picture}(240,200)(0,0)
\Photon(0,150)(25,150){2}{5}
\Text(5,140)[bl]{$v$}
\ArrowLine(25,150)(60,175)
\Text(35,170)[bl]{$f_1$}
\ArrowLine(60,125)(25,150)
\Text(35,130)[tl]{$f_2$}
\ArrowLine(60,175)(60,125)
\Text(65,150)[bl]{$f_3$}
\DashArrowLine(60,175)(85,175){2}
\Text(70,170)[tl]{$s_a$}
\DashArrowLine(85,125)(60,125){2}
\Text(70,120)[tl]{$s_b$}
\Text(55,105)[bl]{$(a)$}
\ArrowLine(65,180)(80,180)
\Text(75,185)[bl]{$k_1$}
\ArrowLine(65,130)(80,130)
\Text(75,135)[bl]{$k_2$}
\Photon(120,150)(145,150){2}{5}
\Text(125,140)[bl]{$v$}
\ArrowLine(180,175)(145,150)
\Text(155,170)[bl]{$f_1$}
\ArrowLine(145,150)(180,125)
\Text(155,130)[tl]{$f_2$}
\ArrowLine(180,125)(180,175)
\Text(185,150)[bl]{$f_3$}
\DashArrowLine(180,175)(205,175){2}
\Text(190,170)[tl]{$s_a$}
\DashArrowLine(205,125)(180,125){2}
\Text(190,120)[tl]{$s_b$}
\Text(175,105)[bl]{$(b)$}
\Photon(0,50)(25,50){2}{5}
\Text(5,40)[bl]{$v$}
\DashArrowLine(25,50)(60,75){2}
\Text(35,70)[bl]{$s_1$}
\DashArrowLine(60,25)(25,50){2}
\Text(35,30)[tl]{$s_2$}
\DashArrowLine(60,75)(60,25){2}
\Text(65,50)[bl]{$s_3$}
\DashArrowLine(60,75)(85,75){2}
\Text(70,70)[tl]{$s_a$}
\DashArrowLine(85,25)(60,25){2}
\Text(70,20)[tl]{$s_b$}
\Text(55,5)[bl]{$(c)$}
\Photon(120,50)(145,50){2}{5}
\Text(125,40)[bl]{$v$}
\DashArrowLine(180,75)(145,50){2}
\Text(155,70)[bl]{$s_1$}
\DashArrowLine(145,50)(180,25){2}
\Text(155,30)[tl]{$s_2$}
\DashArrowLine(180,25)(180,75){2}
\Text(185,50)[bl]{$s_3$}
\DashArrowLine(180,75)(205,75){2}
\Text(190,70)[tl]{$s_a$}
\DashArrowLine(205,25)(180,25){2}
\Text(190,20)[tl]{$s_b$}
\Text(175,5)[bl]{$(d)$}
\Text(105,0)[bl]{$v-s-s$}
\end{picture}
\end{center}
\caption[]{Gauge boson$-$Higgs boson$-$Higgs boson vertex ($vs_as_b$).}
\label{vss}
\end{figure}
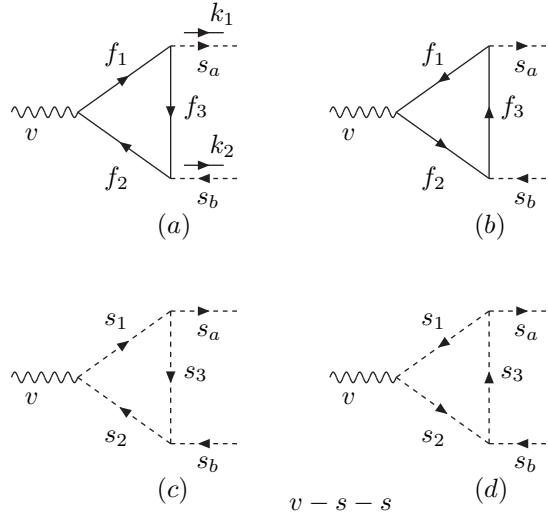
The unrenormalized gauge boson$-$Higgs boson$-$Higgs boson
($v s_a s_b = \gamma h G, \gamma h A, Z h G, ZhA$)
vertex shown in {\bf Fig}.~\ref{vss} ($a$)-($d$) are
\begin{eqnarray}
\Gamma^{v s_a s_b} &=& \Gamma_1 (k_1 - k_2)_\mu + \Gamma_2 (k_1 + k_2)_\mu
\end{eqnarray}
where $\Gamma_{1,2}$ is the summation from $\Gamma_{i(a)}$ to $\Gamma_{i(d)}$,
$\Gamma_i = \Gamma_{i(a)} + \Gamma_{i(b)} + \dots + \Gamma_{i(d)}$,
where $\Gamma_{i(a)}$ and $\Gamma_{i(c)}$ are expressed as follow,
\begin{eqnarray}
\Gamma_{1(a)} &=& \frac{1}{24\pi^2} [-3 (G^{R v}_{\bar{f}_1 f_2}
 G^{L s_a}_{\bar{f}_3 f_1} G^{R s_b}_{\bar{f}_2 f_3} +
 G^{L v}_{\bar{f}_1 f_2} G^{R s_a}_{\bar{f}_3 f_1}
 G^{L s_b}_{\bar{f}_2 f_3}) m_{f_1} m_{f_2} (C_1 + C_2) -
 3 ((G^{R v}_{\bar{f}_1 f_2} G^{L s_a}_{\bar{f}_3 f_1}
 G^{L s_b}_{\bar{f}_2 f_3} + G^{L v}_{\bar{f}_1 f_2}
 G^{R s_a}_{\bar{f}_3 f_1} G^{R s_b}_{\bar{f}_2 f_3}) \nonumber\\
&& m_{f_1} + (G^{L v}_{\bar{f}_1 f_2} G^{L s_a}_{\bar{f}_3 f_1}
 G^{L s_b}_{\bar{f}_2 f_3} + G^{R v}_{\bar{f}_1 f_2}
 G^{R s_a}_{\bar{f}_3 f_1} G^{R s_b}_{\bar{f}_2 f_3}) m_{f_2}) m_{f_3}
 (C_0 + C_1 + C_2)- \nonumber\\
&& (G^{L v}_{\bar{f}_1 f_2} G^{L s_a}_{\bar{f}_3 f_1}
 G^{R s_b}_{\bar{f}_2 f_3} + G^{R v}_{\bar{f}_1 f_2}
 G^{R s_a}_{\bar{f}_3 f_1} G^{L s_b}_{\bar{f}_2 f_3})
 (3 (C_2 + 2 C_{22} + 2 C_{222}) k_1^2 + 2 (12 C_{00} + 9 C_{001} +
 9 C_{002}) - \nonumber\\
&& 6 (C_{112} + 2 C_{12} + C_{222}) k_1 \cdot k_2 + 3 (C_1 + 2 C_{11} +
 C_{111} + C_{112}) k_2^2)] \nonumber\\
\Gamma_{2(a)} &=& \frac{1}{8\pi^2} [(G^{R v}_{\bar{f}_1 f_2}
 G^{L s_a}_{\bar{f}_3 f_1} G^{R s_b}_{\bar{f}_2 f_3} +
 G^{L v}_{\bar{f}_1 f_2} G^{R s_a}_{\bar{f}_3 f_1}
 G^{L s_b}_{\bar{f}_2 f_3}) m_{f_1} m_{f_2} (C_1 - C_2) +
 (G^{R v}_{\bar{f}_1 f_2} G^{L s_a}_{\bar{f}_3 f_1}
 G^{L s_b}_{\bar{f}_2 f_3} + G^{L v}_{\bar{f}_1 f_2}
 G^{R s_a}_{\bar{f}_3 f_1} G^{R s_b}_{\bar{f}_2 f_3}) \nonumber\\
&& m_{f_1} m_{f_3} (C_0 + C_1 - C_2) -
 (G^{L v}_{\bar{f}_1 f_2} G^{L s_a}_{\bar{f}_3 f_1}
 G^{L s_b}_{\bar{f}_2 f_3} + G^{R v}_{\bar{f}_1 f_2}
 G^{R s_a}_{\bar{f}_3 f_1} G^{R s_b}_{\bar{f}_2 f_3}) m_{f_2} m_{f_3}
 (C_0 - C_1 + C_2) - \nonumber\\
&& (G^{L v}_{\bar{f}_1 f_2} G^{L s_a}_{\bar{f}_3 f_1}
 G^{R s_b}_{\bar{f}_2 f_3} + G^{R v}_{\bar{f}_1 f_2}
 G^{R s_a}_{\bar{f}_3 f_1} G^{L s_b}_{\bar{f}_2 f_3})
 (-6 C_{001} + 6 C_{002} - C_2 k_1^2 + 2 (C_{112} - C_{222}) k_1 \cdot k_2 +
\nonumber\\
&&  (C_1 - C_{111} + C_{112}) k_2^2)] \nonumber\\
\Gamma_{1(c)} &=& \frac{1}{8\pi^2} G^{s_a}_{s_3 s_1} G^{s_b}_{s_2 s_3}
 G^{v}_{s_1 s_2} (C_0 + C_1 + C_2) \nonumber\\
\Gamma_{2(c)} &=& \frac{1}{8\pi^2} G^{s_a}_{s_3 s_1} G^{s_b}_{s_2 s_3}
 G^{v}_{s_1 s_2} (-C_1 + C_2)
\end{eqnarray}
with arguments of $C$ functions as
$C(k_2^2, (k_1+k_2)^2, k_1^2, m_{f_3}^2, m_{f_2}^2, m_{f_1}^2)$
and $C(k_2^2, (k_1+k_2)^2, k_1^2, m_{s_3}^2, m_{s_2}^2, m_{s_1}^2)$
for {\bf Fig}.~\ref{svv} ($a$) and ($c$),
and we get the result of {\bf Fig}.~\ref{vss} ($b$) and ($d$) with
the exchange of $s_a \leftrightarrow s_b$ and $f,s_1 \leftrightarrow f,s_2$
from the one of {\bf Fig}.~\ref{vss} ($a$) and ($c$).
For the vertex $\gamma h G$, $\gamma h A$, $Z h G$, $ZhA$,
the virtual fields propogated in the loops are as follow,
$f_1 f_2 f_3 = ttt, bbb$,
$s_1 s_2 s_3 = \tilde{t}_\alpha \tilde{t}_\beta \tilde{t}_\gamma,
\tilde{b}_\alpha \tilde{b}_\beta \tilde{b}_\gamma$.

\begin{figure}
\begin{center}
\begin{picture}(360,200)(0,0)
\ArrowLine(5,170)(20,150)
\ArrowLine(20,150)(5,130)
\Photon(20,150)(40,150){2}{5}
\ArrowLine(40,150)(60,170)
\ArrowLine(60,170)(95,170)
\ArrowLine(60,130)(40,150)
\ArrowLine(95,130)(60,130)
\DashArrowLine(60,170)(80,150){2}
\DashArrowLine(80,150)(60,130){2}
\DashLine(80,150)(95,150){2}
\ArrowLine(125,170)(140,150)
\ArrowLine(140,150)(125,130)
\Photon(140,150)(160,150){2}{5}
\DashArrowLine(180,170)(160,150){2}
\ArrowLine(180,170)(215,170)
\DashArrowLine(160,150)(180,130){2}
\ArrowLine(215,130)(180,130)
\ArrowLine(200,150)(180,170)
\ArrowLine(180,130)(200,150)
\DashLine(200,150)(215,150){2}
\ArrowLine(245,170)(260,150)
\ArrowLine(260,150)(245,130)
\Photon(260,150)(280,150){2}{5}
\ArrowLine(280,150)(300,170)
\DashLine(300,170)(335,170){2}
\ArrowLine(300,130)(280,150)
\ArrowLine(335,130)(300,130)
\ArrowLine(300,170)(320,150)
\DashArrowLine(320,150)(300,130){2}
\ArrowLine(320,150)(335,150)
\ArrowLine(5,70)(20,50)
\ArrowLine(20,50)(5,30)
\Photon(20,50)(40,50){2}{5}
\ArrowLine(40,50)(60,70)
\ArrowLine(60,70)(95,70)
\ArrowLine(60,30)(40,50)
\DashLine(95,30)(60,30){2}
\DashArrowLine(60,70)(80,50){2}
\ArrowLine(80,50)(60,30)
\ArrowLine(95,50)(80,50)
\ArrowLine(125,70)(140,50)
\ArrowLine(140,50)(125,30)
\Photon(140,50)(160,50){2}{5}
\DashArrowLine(180,70)(160,50){2}
\DashLine(180,70)(215,70){2}
\DashArrowLine(160,50)(180,30){2}
\ArrowLine(215,30)(180,30)
\DashArrowLine(200,50)(180,70){2}
\ArrowLine(180,30)(200,50)
\ArrowLine(200,50)(215,50)
\ArrowLine(245,70)(260,50)
\ArrowLine(260,50)(245,30)
\Photon(260,50)(280,50){2}{5}
\DashArrowLine(300,70)(280,50){2}
\ArrowLine(300,70)(335,70)
\DashArrowLine(280,50)(300,30){2}
\DashLine(335,30)(300,30){2}
\ArrowLine(320,50)(300,70)
\DashArrowLine(300,30)(320,50){2}
\ArrowLine(335,50)(320,50)
\end{picture}
\end{center}
\caption[]{Box diagrams.}
\label{box}
\end{figure}
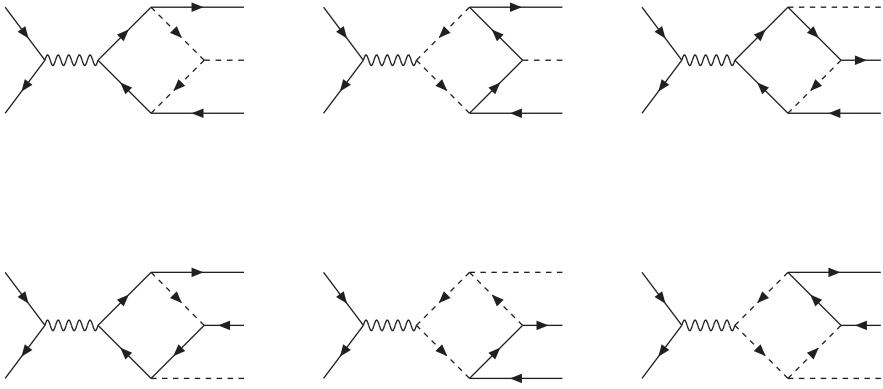

The analytic expressions of box diagrams in {\bf Fig}.~\ref{box}
are too length to present here,
while we keep them in our numerical calculation.

\newpage
\begin{figure}
\begin{center}
\epsfxsize=16cm
\epsfysize=8cm
\epsffile{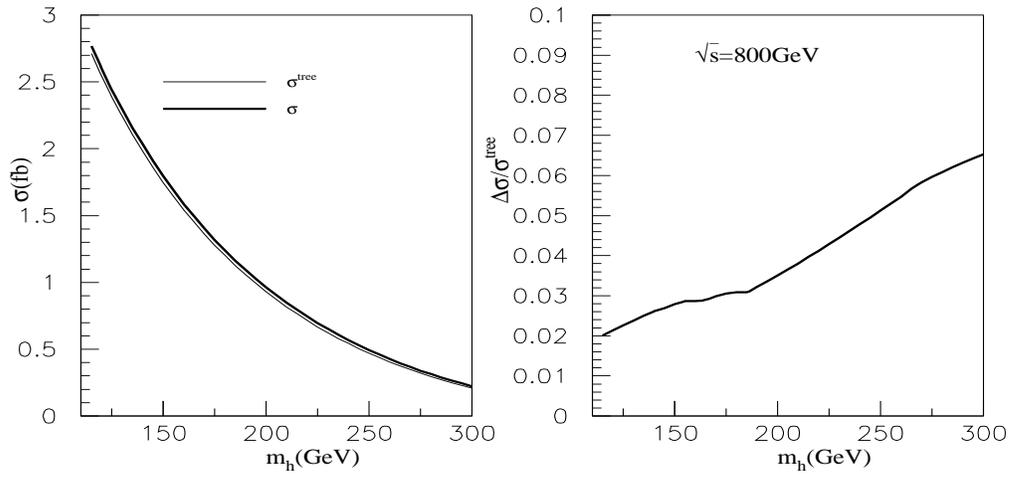}
\end{center}
\caption{$\sigma$(left) and
corresponding $\Delta\sigma/\sigma^{\rm tree}$ (right)
as a function of $m_h$ with $\sqrt{s} = 800$GeV in the SM.
}
\label{smfig}
\end{figure}

\begin{figure}
\begin{center}
\epsfxsize=16cm
\epsfysize=16cm
\epsffile{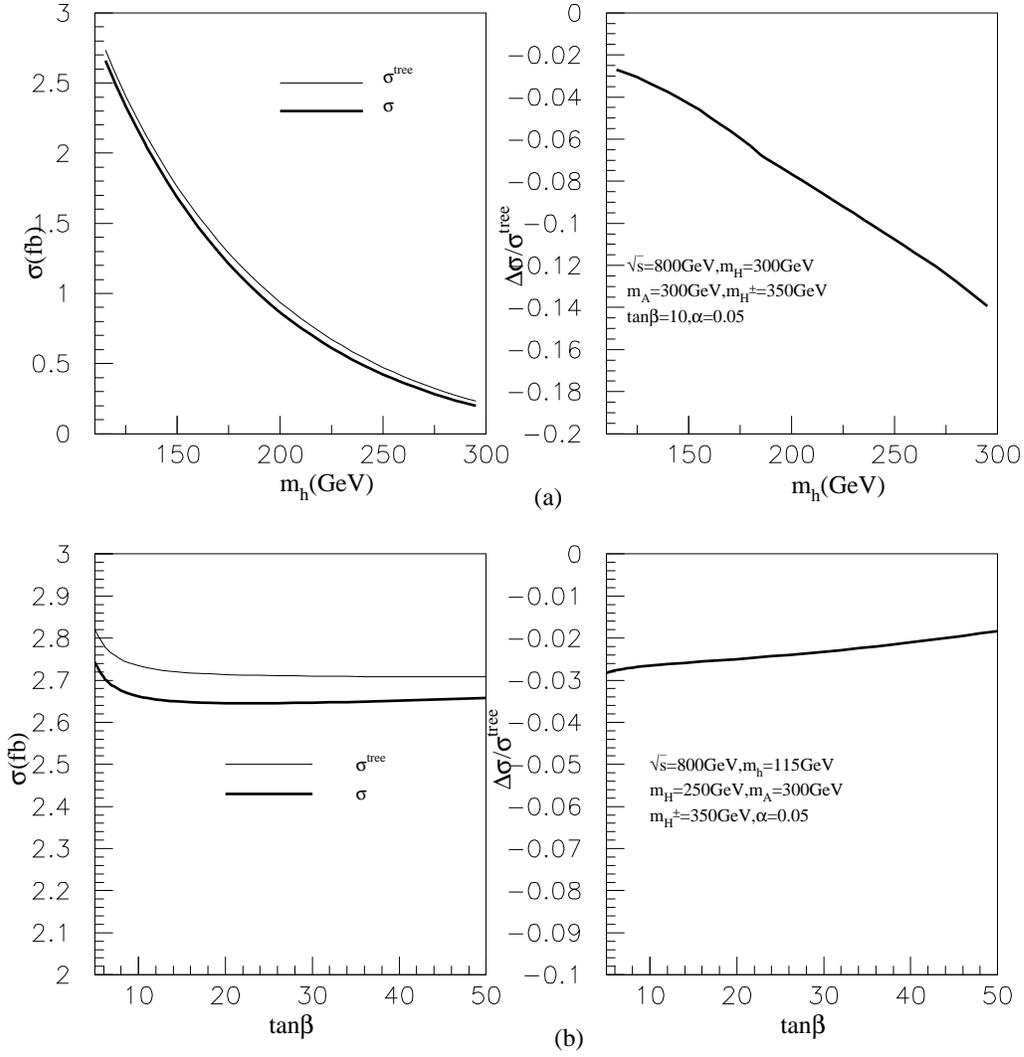}
\end{center}
\caption{$\sigma$(left) and
corresponding $\Delta\sigma/\sigma^{\rm tree}$ (right)
as a function of $m_h$ and $\tan\beta$
corresponding respectively to ($a$), and ($b$)
in 2HDM with $\sqrt{s}=800$GeV.
}
\label{thdmfig}
\end{figure}

\begin{figure}
\begin{center}
\epsfxsize=16cm
\epsfysize=16cm
\epsffile{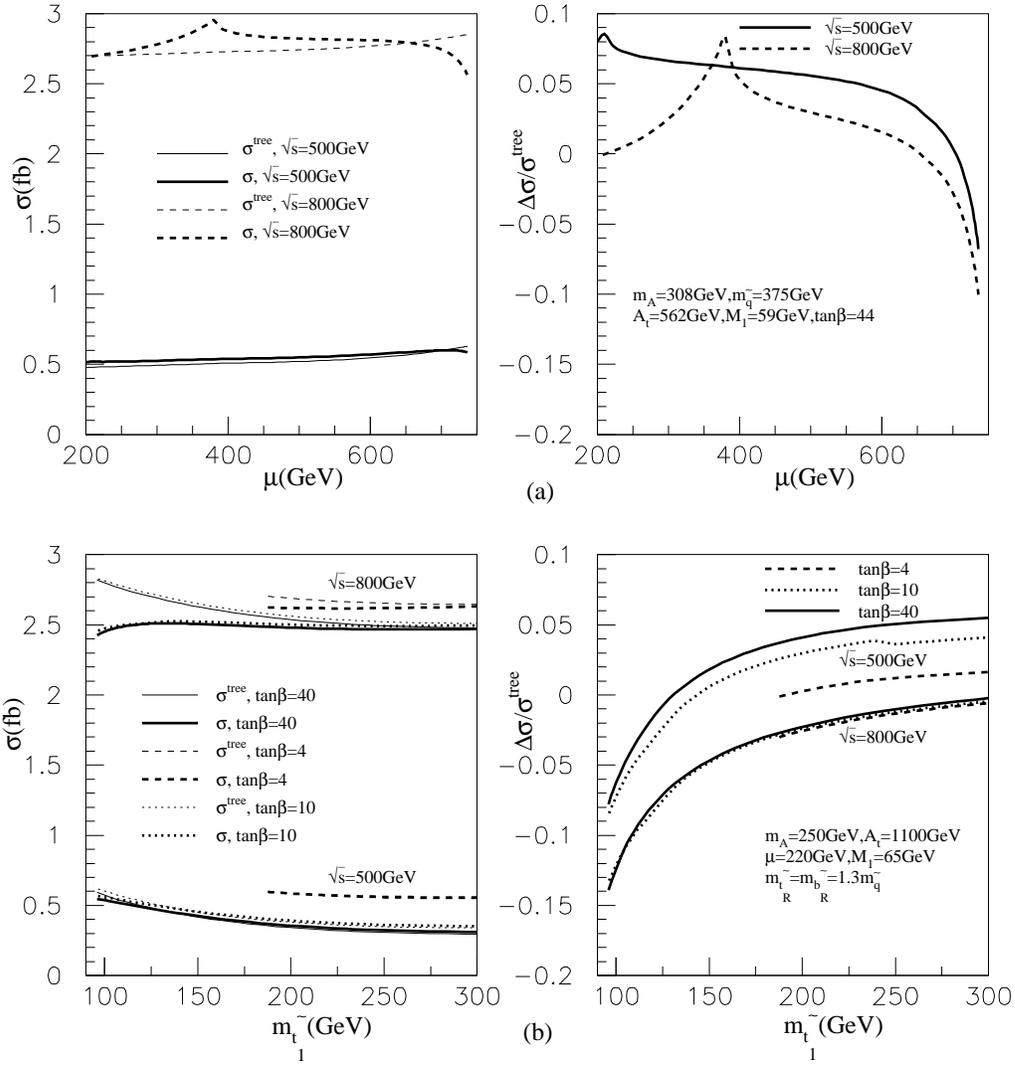}
\end{center}
\caption{$\sigma$(left) and
corresponding $\Delta\sigma/\sigma^{\rm tree}$ (right)
as a function of $\mu$
and the lightest scalar top quark mass $m_{\tilde{t}_1}$
corresponding respectively to ($a$) and ($b$)
in MSSM with $\sqrt{s}=500, 800$GeV.
}
\label{mssmfig}
\end{figure}

\begin{figure}
\begin{center}
\epsfxsize=16cm
\epsfysize=8cm
\epsffile{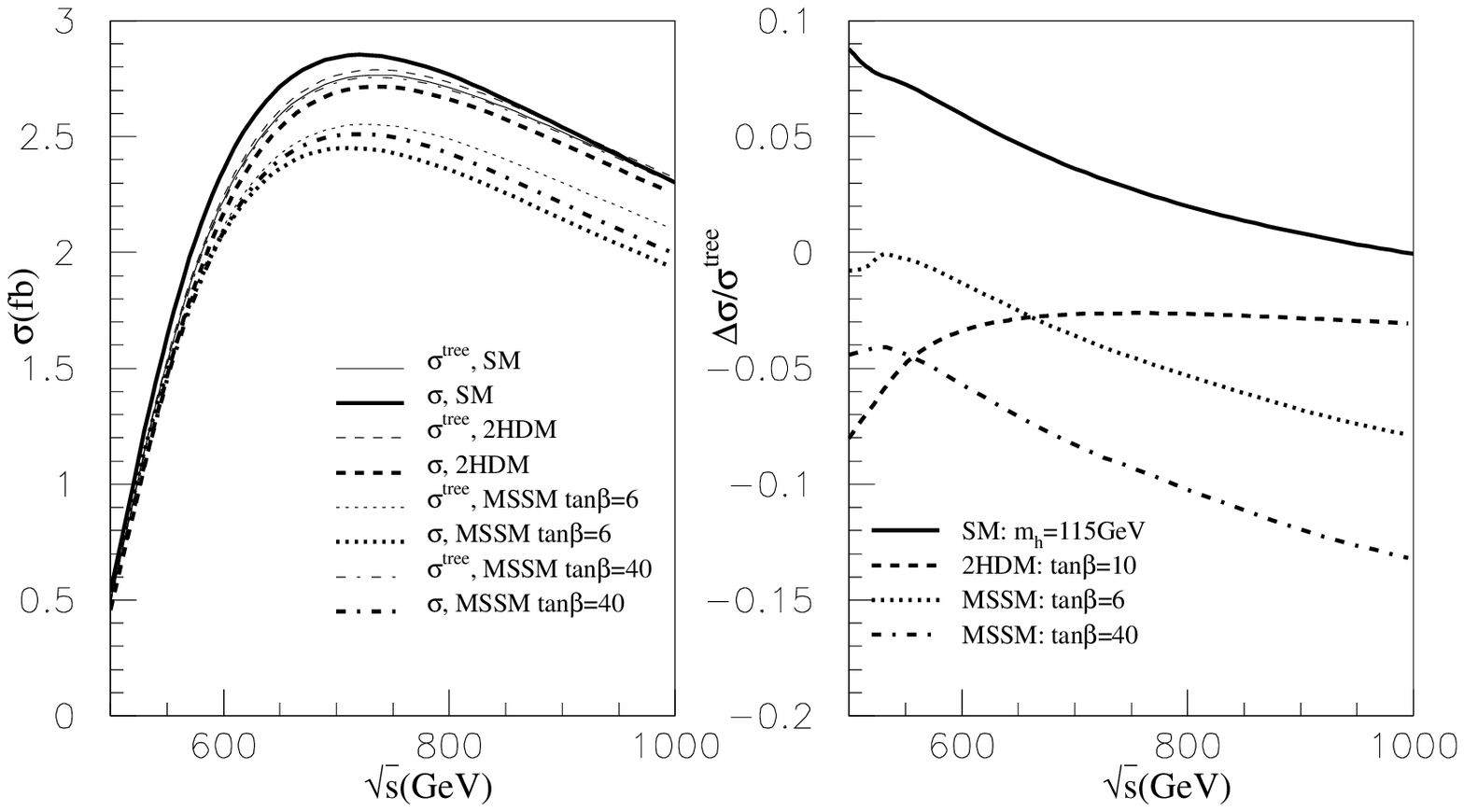}
\end{center}
\caption{$\sigma$(left) and
corresponding $\Delta\sigma/\sigma^{\rm tree}$ (right)
as a function of $\sqrt{s}$ in the SM, 2HDM and MSSM.
The other 2HDM parameters are $m_h=115$GeV, $m_H=250$GeV, $m_A=300$GeV,
$m_{H^\pm}=350$GeV, and $\alpha=0.05$.
And the other MSSM parameters are $M_A=160$GeV, $A_t=1$TeV,
$\mu=220$GeV, $M_1=70$GeV
and $m_{\tilde{t}_R}=m_{\tilde{b}_R}=1.5 m_{\tilde{q}}$.
}
\label{sqrts}
\end{figure}


\begin{thebibliography}{99}
\bibitem{higgs}P.~W.~Higgs, Phys. Lett.  {\bf 12}  (1964) 132;
Phys. Rev. Lett. {\bf 13} (1964) 508; Phys. Rev.  {\bf 145} (1966) 1156;
F.~Englert and R.~Brout, Phys. Rev. Lett. {\bf 13} (1964) 321;
G.~S.~Guralnik, C.~R.~Hagen and T.~W.~Kibble,
Phys. Rev. Lett. {\bf 13} (1964) 585;
T.~W.~Kibble, Phys. Rev. {\bf 155} (1967) 1554.

\bibitem{sm}S.~L.~Glashow, Nucl. Phys. {\bf 22} (1961) 508;
S.~Weinberg, Phys. Rev. Lett. {\bf 19} (1967) 1264;
A.~Salam, in {bf Proceedings of the Eighth Nobel Symposium},
edited by N.~Svartholm (Almqvist and Wiksell,
Stockholm, 1968; Wiley, New York, 1978), p367.

\bibitem{hierarchy}S.~Weinberg, Phys. Rev. {\bf D13} (1976) 974;
Phys. Rev. {\bf D19} (1979) 1277;
L.~Susskind, Phys. Rev. {\bf D20} (1979) 2619;
G.~'t Hooft, in {\bf Recent developments in gauge theories},
Proceedings of the NATO Advanced Summer Institute, Cargese 1979,
ed. G.~'t Hooft et. al. (Plenum, New York 1980).

\bibitem{higgshunter}J.~F.~Gunion, H.~E.~Haber, G.~L.~Kane and S.~Dawson,
The Higgs Hunter's Guide, Addison-Wesley, 1990.

\bibitem{topfound}CDF Collaboration, F.~Abe, et. al.,
Phys. Rev. Lett. {\bf 74} (1995) 2626;
D$0$ Collaboration, F.~Abachi, et. al., Phys. Rev. Lett. {\bf 74} (1995) 2632.

\bibitem{lep2higgssm}LEP Higgs Working Group, arXiv:hep-ex/0107029.

\bibitem{lepewwg02}LEP Electroweak Working Group, LEPEWWG/2002-01, May 8, 2002.

\bibitem{lep2higgsmssm}LEP Higgs Working Group, arXiv:hep-ex/0107030,
arXiv:hep-ex/0107031.

\bibitem{sopczak}A.~Sopczak, arXiv:hep-ph/0112082, arXiv:hep-ph/0112086.

\bibitem{hmax}Y.~Okada, M.~Yamaguchi and T.~Yanagida,
Prog. Theor. Phys. {\bf 85} (1991)1;
H.~Haber and R.~Hempfling, Phys. Rev. Lett. {\bf 66} (1991) 1815;
J.~Ellis, G.~Ridolfi and F.~Zwirner, Phys. Lett. {\bf B257} (1991) 83;
J.~Ellis, G.~Ridolfi and F.~Zwirner, Phys. Lett. {\bf B262} (1991) 477;
R.~Barbieri, F.~Caravaglios and M.~Frigeni, Phys. Lett. {\bf B258} (1991) 167;
R.~Hempfling and A.~Hoang, Phys. Lett. {\bf B331} (1994) 99;
J.~A.~Casas, J.~R.~Espinosa, M.~Quiros and A.~Riotto,
Nucl. Phys. {\bf B436} (1995) 3;
M.~Carena, J.~R.~Espinosa, M.~Quiros and C.~Wagner,
Phys. Lett. {\bf B355} (1995) 209;
H.~Haber, R.~Hempfling and A.~Hoang, Z. Phys. {\bf C75} (1997) 539.

\bibitem{tevatrontth}J.~Goldstein, C.~S.~Hill, J.~Incandela, S.~Parke,
D.~Rainwater, D.~Stuart, Phys. Rev. Lett. {\bf 86} (2001) 1694.

\bibitem{lhctth}E.~Richter-Was and M.~Sapinski,
Acta Phys. Polon. {\bf B30} (1999) 1001;
V.~Drollinger, T.~Muller, D.~Denegri, arXiv:hep-ph/0111312;
F.~Maltoni, D.~Rainwater, S.~Willenbrock, Phys. Rev. {\bf D66} (2002) 034022;
A.~Belyaev and L.~Reina, JHEP {\bf 0208} (2002) 041;
T.~Abe {\it et al.}, arXiv:hep-ex/0106056.

\bibitem{lhcyukawa}For a review, see M.~Beneke {\it et al.},
arXiv:hep-ph/0003033.

\bibitem{hadronqcdzerwas}W.~Beenakker, S.~Dittmaier, M.~Kramer,
B.~Plumper, M.~Spira and P.~M.~Zerwas, Phys. Rev. Lett. {\bf 87} (2001) 201805;
Nucl. Phys. {\bf B653} (2003) 151.

\bibitem{hadronqcddawson}L.~Reina and S.~Dawson,
Phys. Rev. Lett. {\bf 87} (2001) 201804;
S.~Dawson, C.~Jackson, L.~H.~Orr, L.~Reina, D.~Wackeroth, arXiv:hep-ph/0305087.

\bibitem{lctth}J.~F.~Gunion, B.~Grzadkowski and X.~-H.~He,
Phys. Rev. Lett. {\bf 77} (1996) 5172;
S.~Dawson and L.~Reina, Phys. Rev. {\bf D57} (1998) 5851;
Phys. Rev. {\bf D59} (1999) 054012;
S.~Dawson and L.~Reina, Phys. Rev. {\bf D60} (1999) 015003;
A.~Juste and G.~Merino, arXiv:hep-ph/9910301;
H.~Baer, S.~Dawson, and L.~Reina, Phys. Rev. {\bf D61} (2000) 013002;
and J.~F.~Gunion, H.~E.~Haber, R.~V.~Kooten,
arXiv:hep-ph/0301023 for a recent review.

\bibitem{lcqcd}S.~Dittmaier, M.~Kramer, Y.~Liao, M.~Spira and P.~M.~Zerwas,
Phys. Lett. {\bf B441} (1998) 383;
S.~Dawson, and L.~Reina, Phys. Rev. {\bf D59} (1999) 054012;
S.~Dawson, and L.~Reina, Phys. Rev. {\bf D60} (1999) 015003;
S.~Dittmaier, M.~Kramer, Y.~Liao, M.~Spira and P.~M.~Zerwas,
Phys. Lett. {\bf B478} (2000) 247.

\bibitem{treetth}K.~J.~F.~Gaemers and G.~J.~Gounaris,
Phys. Lett. {\bf B77} (1978) 379;
A.~Djouadi, J.~Kalinowski and P.~M.~Zerwas, Zeit. Phys. {\bf C54} (1992) 255.

\bibitem{shzhu}S.~-H.~Zhu, arXiv:hep-ph/0212273.

\bibitem{feyncalc}G.~Passarino and M.~Veltman,
Nucl. Phys, {\bf B160} (1979) 151;
R.~Mertig, M.~Bohm, and A.~Denner, Comp.  Phys. Comm. {\bf 64} (1991) 345.

\bibitem{onshell}S.~Sirlin, Phys. Rev. {\bf D22} (1980) 971;
W.~J.~Marciano and A.~Sirlin, Phys. Rev. {\bf D22} (1980) 2695;
Phys. Rev. {\bf D31} (1985) 213(E);
A.~Sirlin and W.~J.~Marciano, Nucl. Phys. {\bf B189} (1981) 442;
K.~I.~Aoki et al., Prog. Theor. Phys. Suppl. {\bf 73} (1982) 1;
M.~Bohm, H.~Spiesberger and W.~Hollik, Forts. Phys. {\bf 34} (1986) 1;
W.~Hollik, Forts. Phys. {\bf 38} (1990) 3.

\bibitem{denner}A.~Denner, Fortsch. Phys. {\bf 41} (1993) 307.

\bibitem{santos}R.~Sandos and A.~Barroso, Phys, Rev. {\bf D56} (1997) 5366.

\bibitem{hollik}J.~Guasch, J.~Sola, W.~Hollik, Phys. Lett. {\bf B437} (1998) 88;
H.~Eberl, S.~Kraml, W.~Majerotto, JHEP {\bf 9905} (1999) 016.

\bibitem{tanb}A.~Mendez and A.~Pomarol, Phys. Lett. {\bf B279} (1992) 98;
L.~G.~Jin, C.~S.~Li, Phys.Rev. {\bf D65} (2002) 035007.

\bibitem{hewett}J.~Hewett, J.~D.~Wells, Phys. Rev. {\bf D55} (1997) 5549.

\bibitem{subhpole}M.~Carena, M.~Quiros and C.~E.~M.~Wagner,
Nucl. Phys. {\bf B461} (1996) 407;
M.~Carena, H.~E.~Haber, S.~Heinemeyer, W.~Hollik, C.~E.~M.~Wagner,
and G.~Weiglein, Nucl. Phys. {\bf B580} (2000) 29.

\bibitem{pdg}K.~Hagiwara {\it et al.}, Particle Data Group,
Phys. Rev. {\bf D66} (2002) 010001.

\bibitem{rho}M.~Drees and K.~Hagiwara, Phys. Rev. {\bf D42} (1990) 1709;
M.~Drees, K.~Hagiwara and A.~Yamada, Phys. Rev. {\bf D45}  (1992) 1725;
P.~Chankowski {\it et al.}, Nucl. Phys.  {\bf B417} (1994) 101;
D.~Garcia and J.~Sol\`a, Mod. Phys. Lett. {\bf A9} (1994) 211;
A.~Djouadi {\it et al.}, Phys. Rev. Lett. {\bf 78} (1997) 3626.

\bibitem{carena}M.~Carena, S.~Heinemeyer, C.~E.~M.~Wagner,
G.~Weiglein, arXiv:hep-ph/9912223.


\bibitem{ma}Y.~You, W.~-G.~Ma, H.~Chen, R.~-Y.~Zhang, Y.~-B.~Sun, H.~-S.~Hou,
arXiv:hep-ph/0306036.

\bibitem{belanger}G.~Belanger, F.~Boudjema, J.~Fujimoto, T.~Ishikawa,
T.~Kaneko, K.~Kato, Y.~Shimizu and Y.~Yasui, arXiv:hep-ph/0307029.

\bibitem{dennertth}A.~Denner, S.~Dittmaier, M.~Roth, M.~M.~Weber,
arXiv:hep-ph/0307193.

\bibitem{haberkane}H.~E.~Haber and G.~L.~Kane, Phys. Rep. {\bf 117} (1985) 75.
\end{thebibliography}
\end{document}